# Finite-size and quark mass effects on the QCD spectrum with two flavors

Claude Bernard,(1) Tom Blum,(2)
Thomas A. DeGrand,(3) Carleton DeTar,(4)
Steven Gottlieb,(5,6) Alex Krasnitz,(5,7) R. L. Sugar,(8) D. Toussaint(2)

(1) Department of Physics, Washington University, St. Louis, MO 63130, USA
(2) Department of Physics, University of Arizona, Tucson, AZ 85721, USA
(3) Physics Department, University of Colorado, Boulder, CO 80309, USA
(4) Physics Department, University of Utah, Salt Lake City, UT 84112, USA
(5) Department of Physics, Indiana University, Bloomington, IN 47405, USA
(6) Department of Physics, Bldg. 510A, Brookhaven National Laboratory, Upton, NY 11973, USA
(7) IPS, RZ F 3, ETH-Zentrum CH-8092 Zurich, SWITZERLAND
(8) Department of Physics, University of California, Santa Barbara, CA 93106, USA

ABSTRACT:

We have carried out spectrum calculations with two flavors of dynamical Kogut-Susskind quarks on four lattice sizes from $8^3 \times 24$ to $16^3 \times 24$ at couplings that correspond to chiral symmetry restoration for a lattice with 6 time slices. We estimate that the linear spatial sizes of the lattices range from 1.8 to 3.6 fm. We find significant finite size effects for all particles between the smallest and largest volume with the larger quark mass that we study, $am_q = 0.025$, where $a$ is the lattice spacing. The nucleon experiences the largest effect of about 6 percent. We also study a lighter quark mass, $am_q = 0.0125$, on the two largest lattices. Effects of the dynamical and valence quark masses on the hadron spectrum are studied both directly, by comparing the two simulations, and by extracting mass derivatives from the correlation functions. We do not find much improvement in the nucleon to rho mass ratio as we decrease the quark mass at this lattice spacing. Finally, we report on an unsuccessful attempt to see effects of the $\rho \to 2\pi$ decay on the $\rho$ mass, and on studies of Wilson and Kogut-Susskind hadron masses with large valence quark masses.

## I. INTRODUCTION AND MOTIVATION

In this paper, we describe a recent lattice QCD spectrum calculation with two flavors of dynamical Kogut-Susskind quarks. As is well known, lattice calculations can qualitatively reproduce the spectrum of low lying hadrons; however, it has not been possible to reproduce directly the nucleon to rho mass ratio[1]. This shortcoming has been independent of whether Kogut-Susskind quarks or Wilson quarks are used and independent of whether the calculation is done in the quenched approximation or with dynamical quarks[2-5]. However, in a recent series of quenched Wilson quark calculations, it has been possible to produce hadron mass ratios in good agreement with the real world when extrapolating in quark mass, volume and lattice spacing[6]. An earlier calculation at $6/g^2 = 6.0$ suggested that this could be done merely by extrapolating in quark mass[7]. Getting the hadron masses correct at one (or a few) percent precision would be a great triumph for lattice QCD and would give one confidence that more detailed quantities such as weak matrix elements[8] and structure functions[9] could be accurately calculated. At this point, it is still not clear exactly what such a calculation would require. That is, lattice calculations are subject to systematic errors arising from a number of sources, and we do not yet have a firm basis for controlling these effects, or for determining which of them are of most crucial importance.

Aside from the numerical parameters related to a particular choice of algorithm, there are at least three parameters that may cause systematic errors: the lattice spacing $a$ (equivalently, the gauge coupling $6/g^2$), the quark mass $m_q$ and the spatial volume $V = (aL)^3$, where $L$ is the number of lattice points in each spatial direction. In addition, if the number of lattice points in the time direction is too small, one may not be able to see the asymptotic falloff of the propagators from which the masses are calculated. In order to carry out a simulation, two discrete choices must also be made: whether or not to use the quenched approximation and whether to use Wilson or Kogut-Susskind quarks. If QCD is the correct theory of the strong interaction and we choose dynamical quarks, then as $a \to 0$, $V \to \infty$ and $m_q$ approaches its physical limit, we will get the correct hadron masses. Unfortunately, each of these limits makes the calculation more difficult. With current computers, it is impossible to approach all of these limits in a single calculation. In this calculation, we attempt to make progress on large volume and small quark mass by working at moderately strong coupling. We have made this choice for the following reasons:

High Energy Monte Carlo Grand Challenge (HEMCGC) calculations with Kogut-Susskind quarks[2] at $6/g^2 = 5.6$ hinted at some improvement in the nucleon to rho mass ratio over stronger coupling calculations[10] done at $\approx 5.4$; however, the errors in the stronger coupling results were quite large, and a considerable improvement in precision was necessary to make a sensible comparison.

We were also very strongly motivated by the striking finite-size effects seen by Fukugita, Mino, Okawa and Ukawa[11]. Their calculation at $6/g^2 = 5.7$ on lattices from $L = 8$ to 20, showed the pion dropping in mass by almost a factor of two from the smallest to the largest size. In fact, the masses for the nucleon and rho are still clearly dropping as $L$ grows from 16 to 20. This was not surprising to us in view of the HEMCGC work at $6/g^2 = 5.6$ with $L = 12$ and 16, where it was seen that the nucleon mass had dropped 10% when the volume was increased. ($L = 16$ at $6/g^2 = 5.6$ is about the same physical size as $L = 21$ at $6/g^2 = 5.7$, where we use the rho mass at zero quark mass to estimate the scale.) Since no larger volume had been explored, there was no direct evidence that $L = 16$ was sufficient to control the finite size effects even with the larger lattice spacing corresponding to $6/g^2 = 5.6$. Nevertheless, the calculation of Fukugita *et al.* with its range of lattice sizes emphasizes how little we know from simulations about the physical box size required to obtain accurate hadron masses.

It is then clear that if our desire is to study large physical volumes and find a range of volumes for which there are small but measurable finite size effects, we will have to work at stronger coupling. We wish to find such a region because we want to demonstrate control over the finite size effects at the level of a percent or so. This also means that we must have very high statistics. Clearly, one must decide whether it is more important to go to weaker coupling with lower statistics or to collect higher statistics at stronger coupling with similar physical volumes. We have chosen the latter.

The need for high statistics simulations is also a mark of the maturity of this field. In the heavy quark mass limit, the nucleon to rho mass ratio is 3/2, the ratio of the number of quarks. In the real world, the ratio is about 1.2. With one FLOP, we can get within 30% of the correct value. Thus, if a simulation can only get the ratio to 10% accuracy, in all probability it will be within 3 standard deviations ($\sigma$) of the experimental value. In fact, we already have many such calculations, and at this point what is needed is a careful comparison of different simulations to get quantitative control over the systematic errors. Interesting physical questions require a higher degree of precision than distinguishing between 1.5 and 1.2. For example, comparing the Edinburgh plots for Wilson and Kogut-Susskind quarks clearly requires very high precision in the masses because the differences between the ratios are small.

It is also important to explore the light quark mass region. There have been many calculations with $m_\pi/m_\rho$ greater than 0.5. However, the lighter quark mass region is not nearly so well studied. In this region, we expect the nucleon to rho mass ratio to be decreasing toward its physical value. In dynamical quark calculations, we also expect that we will have to deal with the effect of the rho becoming unstable[12-17]. This may be one of the largest sources of differences between quenched and dynamical calculations, so it is very important to explore this region directly rather than just extrapolating from heavier masses. Again, it is much more feasible to explore this region with moderately strong coupling than it is to try to do this with weaker couplings. However, it must be noted that at stronger coupling flavor symmetry is broken, so that the number of light pions may be smaller than in nature.

In this moderately strong coupling calculation, we have made the pion mass light enough so that a rho meson with nonzero momentum could decay into two pions. However, we were unable to see the effect of this decay on the mass of the $\rho$.

Having good control of finite-size effects in the moderate coupling region can be of great value for weaker coupling calculations: by determining an adequate physical box size from a moderate coupling calculation, we avoid having to do calculations at weak coupling with a larger physical box size.

Further information about the extrapolation of hadron masses to the real world quark mass can be obtained from the correlation of the hadron propagators with the chiral condensate $\bar{\psi}\psi$. Such correlations can be related to the derivatives of hadron masses with respect to the dynamical quark mass on our lattices. We have computed these derivatives; the results do not give any support to hopes that the spectrum can be extrapolated to the physical spectrum at our large lattice spacing.

The plan of the rest of this paper is as follows. In Sec. II, we briefly discuss our simulation and methodology. Section III details the masses that we have determined for the hadrons in each simulation, including our analysis of the finite-size effects, a discussion of finite-size effects in terms of a very simple picture of nuclear matter, and the rho meson decay. Section IV discusses our calculation of the derivatives of hadron masses with respect to quark mass. Section V gives a comparison of Wilson and Kogut-Susskind Edinburgh plots and discusses their difference in the region where $m_\pi/m_\rho$ is large. Section VI contains our conclusions.

## II. SIMULATION AND METHODOLOGY

Our simulation of two flavor QCD was done using the version of the hybrid molecular dynamics algorithm called the R-algorithm in Ref. 18. The main technical advance over the spectrum calculation first done with that algorithm[10] is the use of a wall source[19] rather than a point source. The other improvements come from using a bigger spatial size and collecting more statistics on a faster computer.

For the $am_q = 0.025$ runs, we use a molecular dynamics step size of 0.02 in the normalization of Ref. 10. For the mass 0.0125, runs we use a step size of 0.01. For the $N_s = 12$ run, we integrated for 0.5 time units between refreshing the momenta. For the $N_s = 16$ run, we integrated for a full time unit. In Table I, we summarize the parameters and lengths of our runs.

In order to obtain the hadron masses, we first fixed the gauge to lattice Coulomb gauge and then calculated the quark propagators. We constructed the hadron propagators with a point sink and a "corner wall" source in which the $(0, 0, 0)$ element of each $2^3$ cube on the selected time slice is set to one. We then used the full covariance matrix of the propagator to carry out a fit[20]. To compute the covariance matrix, we block together propagators from several successive time units of running. With the exception of the nonzero momentum mesons and delta at $am_q = 0.025$, we measured every two units of molecular dynamics time. We used two or more wall sources at time slices spread through the lattice.

## III. HADRON MASSES

We have analyzed the masses of the pion, rho and nucleon in all of our data sets to study the effects of lattice size. In addition, we have looked at some states with nonzero momentum on the $16^3$ lattices. Finally, we have looked at the delta on the largest volume for both masses and the smallest volume with the large mass.

The hadron masses for the different lattice sizes are shown in Tables II and III. Table II contains the zero momentum masses and Table III contains the masses for nonzero momentum along with some relevant zero momentum masses.

### A. Finite Size Effects

For the masses in Tables II we block together propagators from 20 successive time units of running, using two wall sources on each lattice at $t = 0$ and 12. We have examined the autocorrelation of the propagator and find that with few exceptions the autocorrelation is less than 0.1 for a time lag of 10. The pion has larger autocorrelations than other hadron propagators. For the lighter mass and larger volume, we have also tried to estimate the integrated autocorrelation time for the pion propagator at distance 8 from the source. We find that $\tau_{int} = 4.0 \pm 1.3$. Given the above, a block size of 20 time units seems adequate. As an additional check, for the heavier mass, we have compared fits with a block size of 20 with a block size of 10. For the pion, the computed error bars are about 10% bigger with the larger block size. (If the errors grow in proportion to the inverse of the block size, our reported errors for the pion might be about 10% smaller than with infinite block size.) For the rho meson, the large block size has errors 10% or less larger than for the smaller block size. For the nucleon, the cases we compared were within 3% and neither block size was consistently larger. We note that the plaquette has a larger autocorrelation time than the hadron propagators. For the light mass and large volume, we estimate the integrated autocorrelation time for the plaquette to be $19 \pm 4$.

The choice of fit is important, so we detail how we chose the ones we use for our final mass determinations. In general, we are trying to balance two competing effects: we want to get away from the source point to allow excited state contributions to decrease, but we do not want to get too far away as statistical errors become more important far from the source. We fit the propagators from $D_{min}$ to the center of the lattice for all particles except the nucleon. Because of antiperiodic boundary conditions in time, the nucleon propagator should vanish at the center of the lattice so we ignore that distance from the source. For all particles except the pion, we report masses assuming there are two particles of opposite parity propagating in each channel. For the $\pi$ channel, we assume two pseudoscalars. In order to pick the optimal value of $D_{min}$, we have considered the combined confidence level (CCL) of the fits to all four volumes for the heavier quark mass. For small values of $D_{min}$ the confidence levels are very small because there are more than two particles contributing at short distances. To minimize statistical errors, we pick the smallest value of $D_{min}$ for which the CCL is reasonable. In Table V, we show the CCL for several channels for various $D_{min}$. Focusing on the $\rho$ channel, we see that with $D_{min} = 3$ the CCL is infinitesimal, and that it is a maximum with $D_{min} = 4$. The same is true for the $\pi_2$, $\rho_2$ and nucleon channels, and so we report masses from $D_{min} = 4$ fits. We note that for the nucleon, the CCL is quite small due primarily to poorer fits for $N_s = 16$. For example, for $D_{min} = 4$ and 5, the other three fits have a CCL of 0.54 and 0.44, respectively.

For the $\pi$ channel, we see that there are a large number of fits with quite reasonable confidence level. We have previously reported masses for this channel based on single particle fits with $D_{min} = 8$[21]. For the single particle fits, it turns out that the rate of approach to the asymptotic mass is quite dependent upon $N_s$. The one and two particle fits as a function of $D_{min}$ are shown in Fig. 1. For the smallest lattice, $N_s = 8$, we reach a plateau with $D_{min} = 5$. For $N_s = 10$ and 12, we require $D_{min} = 7$ and 8, respectively. For $N_s = 16$, it is not even clear that we see a plateau. As we increase the size of our wall source, we get a stronger coupling to the excited states and, thus, it requires a greater distance from the source to suppress that contribution. For instance, the ratio of the excited state to ground state amplitudes is about $-0.23$ for $N_s = 8$ and steadily increases in magnitude to $-0.32$ for $N_s = 16$. For our table, we choose $D_{min} = 2$ for two reasons. First, larger values of $D_{min}$ may have somewhat larger values of the CCL, but they also have larger error bars. Second, although $D_{min} = 1$ has an even larger CCL than $D_{min} = 2$, for $N_s = 8$ there is a tendency for the pion mass from the two particle fits to decrease as we increase $D_{min}$. We do not see this for other values of $N_s$. The $D_{min} = 2$ fit may be somewhat less than one standard deviation high for $N_s = 8$.

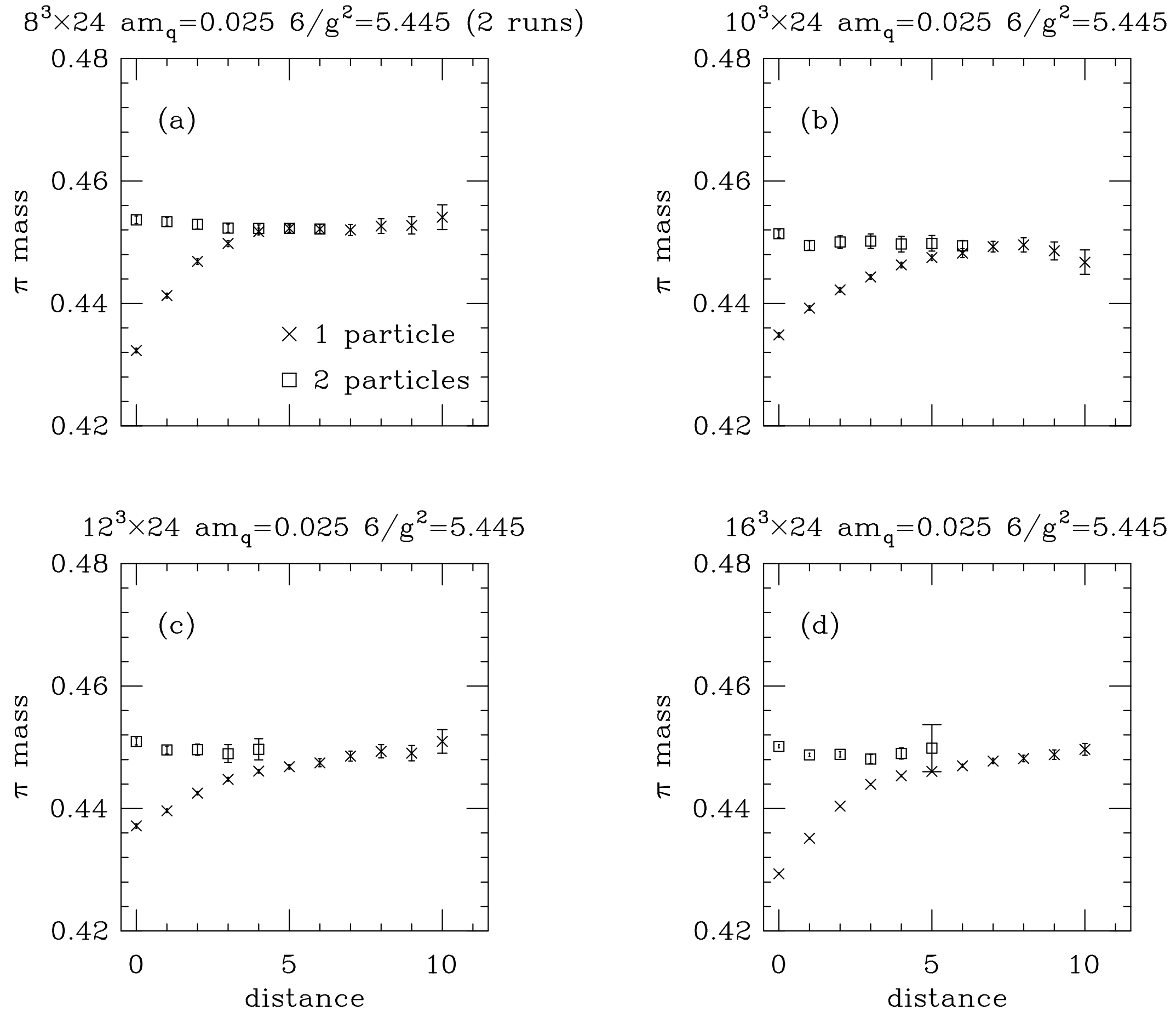

Fig. 1. The pion mass from both one and two particle fits as a function of $D_{min}$. (a) $N_s = 8$, (b) $N_s = 10$, (c) $N_s = 12$, (d) $N_s = 16$. Note that the pion mass as determined by the single particle fits requires larger values of $D_{min}$ to reach its asymptotic value as we increase $N_s$.

For the lighter quark mass $am_q = 0.0125$, we only have two volumes, and there is no reason to use the same values of $D_{min}$ as for $am_q = 0.025$ since the opposite parity and excited state masses and couplings will be different. Looking at the fits for both volumes, we choose the $D_{min} = 3$ fits for all particles.

To summarize, we report masses from two particle fits in each case. For the lighter quark mass, we always use $D_{min} = 3$, but for the heavier quark mass $D_{min}$ depends upon the channel. For the $\pi$ channel, we report masses with $D_{min} = 2$, while for the other channels, we use $D_{min} = 4$. For the nucleon, we omit the center plane of the lattice because of the antiperiodic boundary conditions.

The mass of the pion is determined with the greatest precision. The study of gauge fixed hadron wave functions[22] indicates that the pion has the smallest spatial extent, so we expect the smallest finite size effect for this particle. First, we focus upon the $am_q = 0.025$ results. From $N_s = 8$ to 16, we find the pion mass (in units of the lattice spacing) varies by 0.0041(8)(5.1 $\sigma$). This is a 0.9% effect on the pion mass. In Fig. 2, we plot the pion mass as a function of the spatial size of the lattice. We also have plotted a line indicating the size of a one percent effect. Perhaps all of the finite size effect for the pion is between $N_s = 8$ and 10. That is to say, the values for $N_s = 10$, 12 and 16 are not significantly different given the size of our errors. For the lighter quark mass, we only have results for the two larger lattice sizes, $N_s = 12$ and 16. Here we do not see a significant difference. It would be very valuable to have results for $N_s = 8$ in order to compare the two quark masses over the same range of $N_s$. To convert from $N_s$ to physical volume, we must determine the lattice spacing. If we do this by assuming the rho mass takes its physical value for $am_q = 0.0125$, then $N_s = 8$ (16) corresponds to 1.8 (3.6) fm. (If we were to extrapolate the rho mass to zero quark mass, the smaller box would be 1.73 fm.)

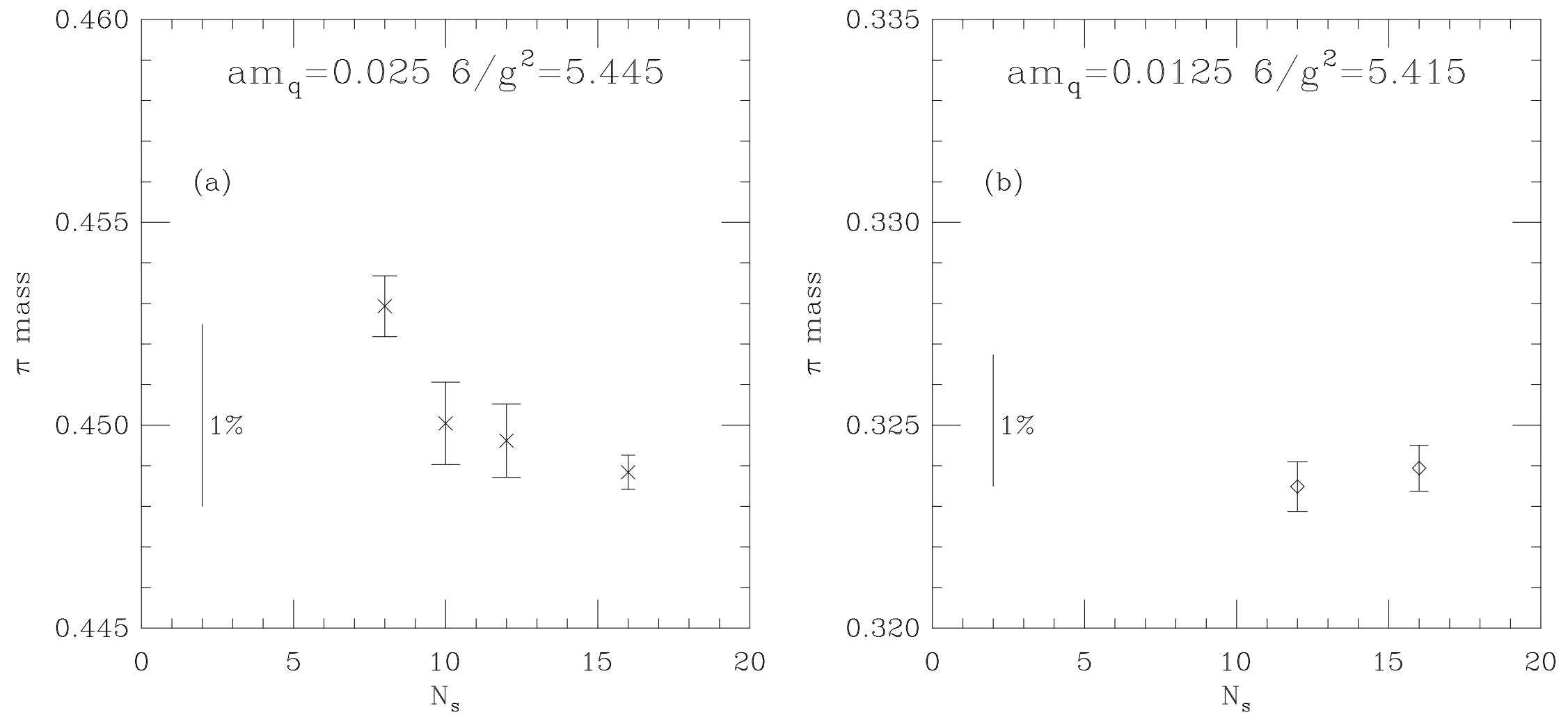

Fig. 2. (a) The pion mass as a function of lattice spatial size for $am_q = 0.025$. A line the height of a 1 percent effect is plotted near the left edge of the graph. (b) Same as (a) except for $am_q = 0.0125$.

The nucleon appears to have the largest finite size effect of the three particles. From $N_s = 8$ to 16, we find the nucleon mass difference is 0.081(23), or a 5.9±1.7% effect. This is much larger than for the pion; however, because the errors are larger, its statistical significance (3.5 $\sigma$) is not as great. The nucleon mass as a function of lattice size is shown in Fig. 3 for both quark masses. In both cases, the nucleon mass is lower for $N_s = 12$ than it is for 16, though not significantly so.

For the rho masses, the difference is 0.031(13) between $N_s = 8$ and 16, or a 3.4% effect which is nonzero by 2.4 $\sigma$. Referring to Fig. 4, we note that the point at $N_s = 12$ is lower than that at 16, as was the case with the nucleon. Once again, however, the difference is not statistically significant. For the lighter quark mass, as we only have results for $N_s = 12$ and 16, we cannot determine a significant finite volume effect; the rho mass is about one

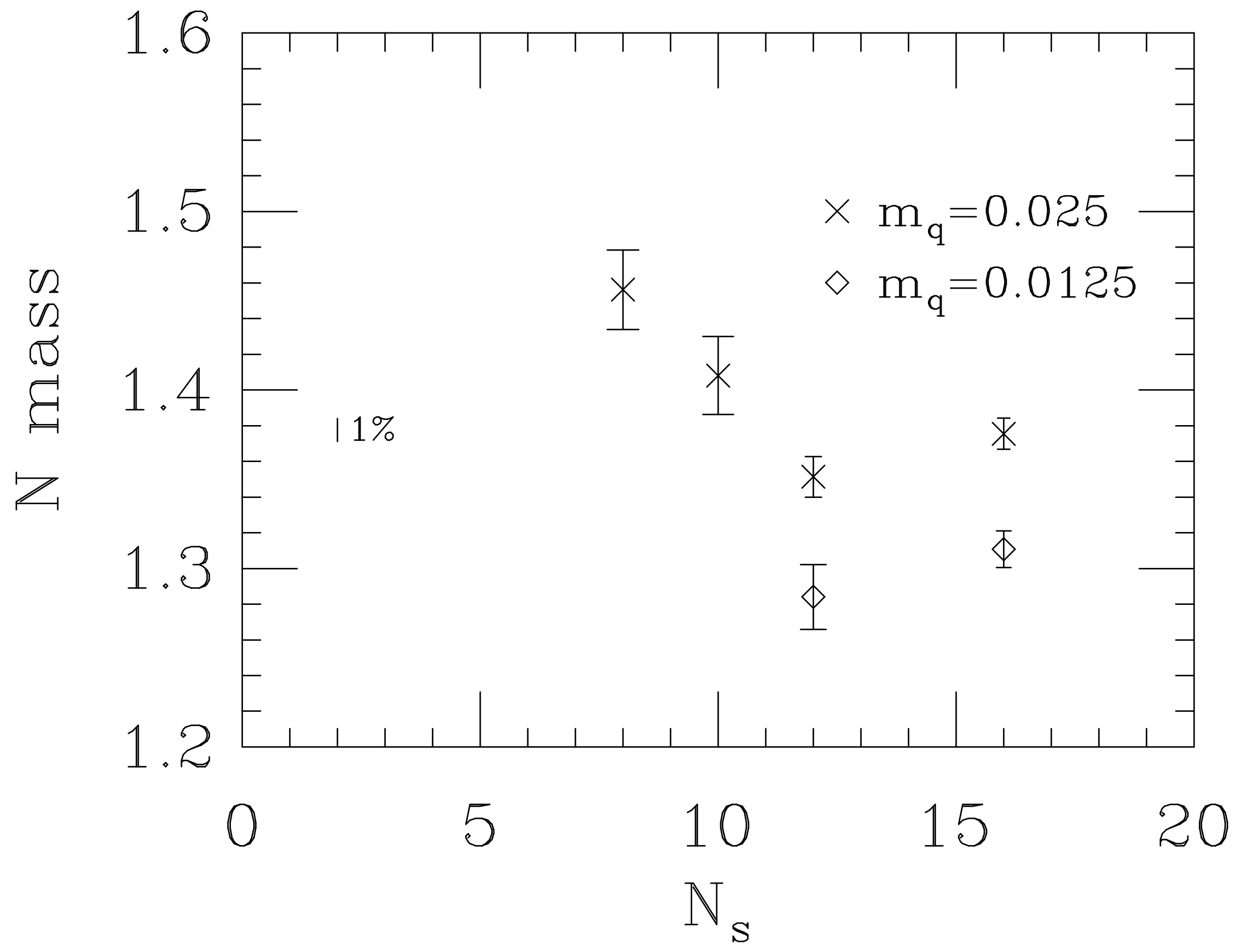

Fig. 3. The nucleon mass as a function of lattice size for both quark masses. The vertical line at the left edge of the graph corresponds to 1 percent error for the heavier nucleon mass.

$\sigma$ heavier for the smaller volume.

We would like to compare our finite size effects with those observed in weaker coupling by preparing a graph of the rho mass as a function of the box size. Because we have different values of $a$, the dimensionless values of $am_\rho$ calculated on the lattice will differ quite a bit. For each coupling, we fit the $\rho$ mass as a function of volume to determine $\rho_\infty$. Then we plot $m_\rho/m_{\rho_\infty}$ versus the physical box size $aN_s$. This requires a knowledge of $a$ which we determine by assuming that $am_{\rho_\infty} = 770$ MeV. It would be possible to refine this by extrapolating the hadron masses in $m_q$ to determine the scale from the rho mass at the point where, say, $m_\pi/m_\rho$ takes its physical value. (In fact, for the weaker coupling results, the extrapolation to zero quark mass has been done and would result in a 16% reduction in the lattice size[23]. For our results, there would be an 8% reduction.) Extrapolating our results to infinite volume using the form $am_\infty + b/V$, following Ref. 23, we find that $am_{\rho_\infty} = 0.911$ at $6/g^2 = 5.445$, and that $am_{\rho_\infty} = 0.409$ at $6/g^2 = 5.7$. In Fig. 5, we show our results with diamonds. The crosses are the results of Ref. 11, and the square is from Ref. 4. In addition to the data, we have plotted two horizontal dotted lines showing an error band of ±2%. We also show a vertical line at a size corresponding to $T_c^{-1}$. It has been known for quite some time from quenched calculations that when the spatial size of the lattice approaches $T_c^{-1}$, there are large effects on the hadron masses. This has been called "spatial deconfinement." Since our simulation parameters correspond to the finite temperature crossover for six time slices, by using a minimum $N_s$ of eight, we avoid such an effect.

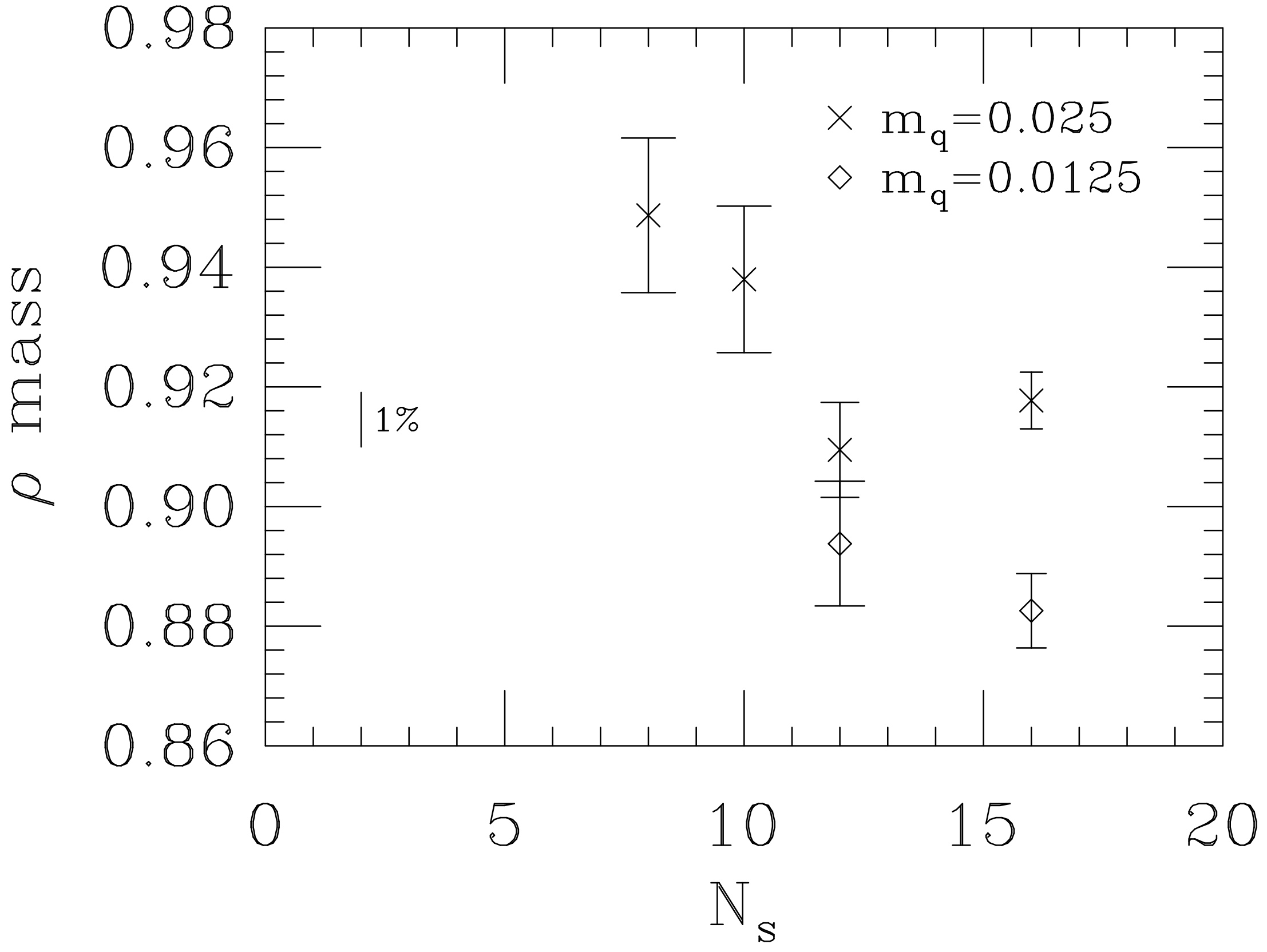

Fig. 4. Same as Fig. 3, but for the rho.

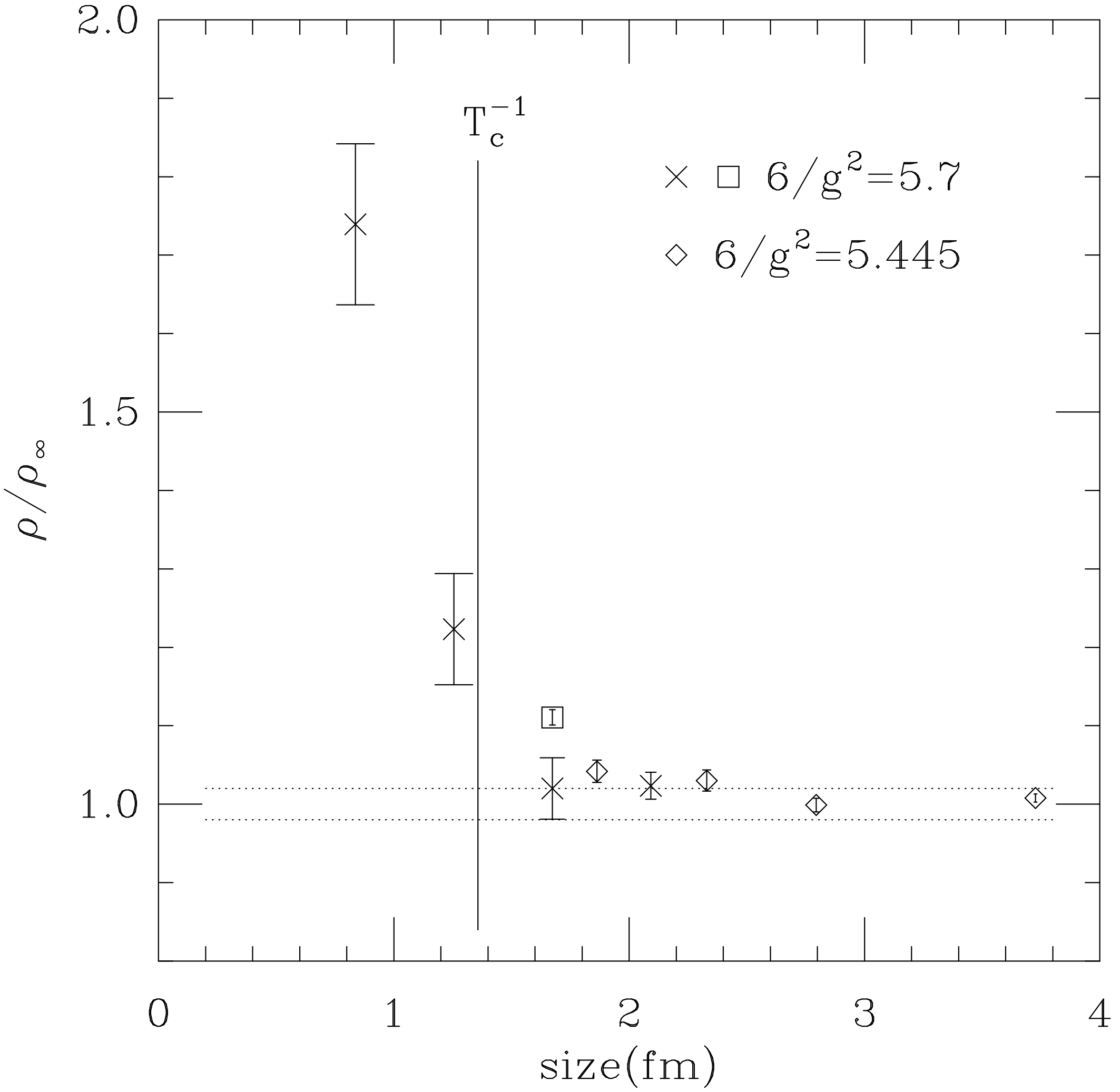

Fig. 5. The $\rho$ mass divided by the infinite volume mass as a function of the linear size of the box. The diamonds are the values from the current calculation. The crosses and square come from weaker coupling results [11,4]. The horizontal lines correspond to a 2% error band. The vertical line shows the spatial size corresponding to $T_c$.

Some time ago, an analytic study of finite volume effects based on pion exchange was carried out by Lüscher[24]. This approach gives the leading terms for large volume. It predicts that the nucleon mass should approach its asymptotic value with a correction roughly like $\exp(-m_\pi L)$. Practical lattice calculations are not done in this regime, and lattice calculations have seen effects much larger than predicted. A power law approach to the large volume limit has been suggested on the basis of the analysis in Ref. 23. That work considers lattice studies where the box length is less than 1.8 fm.

Because of the periodic boundary conditions, we may think of our nucleon in a box as nuclear matter at nonzero density. A lattice calculation done with periodic boundary conditions is a little like finite density nuclear matter because the nucleon "sees" the forces from its periodic images. Of course, the images all move in lockstep unlike real nuclear matter where the nucleons move independently. We make the further simplification of considering a "nuclear crystal" of hexagonal close packed nucleons to estimate the nucleon spacing. A large nucleus is more like a liquid than a crystal, but the difference in density between water and ice, for instance, is not that great.

The density of nuclear matter is 0.16 nucleons per cubic fermi[25]. The density of hexagonal close packed balls is $1/4\sqrt{2}R^3$, where $R$ is the radius of the ball. Solving for the diameter of a nucleon, we get 2.6 fm. Thus, we expect that if the box size is less than 2.6 fm, we are squeezing the nucleons together. At higher density, the energy per nucleon will rise (rapidly if there is a hard core potential). As we decrease the density, we expect the energy per nucleon to approach the nucleon mass from below. If we take our lattice spacing from the rho mass with $am_q = 0.0125$, a box size of 2.6 fm corresponds to 11.5 lattice spacings for the current calculation.

How large might we expect the binding to be in this picture? If we look at the curve of binding energy per nucleon, it is largest for iron at about 9 MeV per nucleon[25]. In a real nucleus, there is Coulomb repulsion while there is none in the lattice calculation. Neglecting the repulsion, nuclear matter models set the binding energy at about 16 MeV. So in this picture, we expect less than a 2% lowering of the nucleon mass from the nuclear attraction. Of course, finite size effects can be much larger for small distances as we squeeze the nucleon, but here we have a higher value than the mass. Another reason that the effect should be smaller on a cubic lattice is that there are only six nearest neighbors, whereas for a hexagonal lattice there are 12.

Despite this nice physical picture, whether we are just seeing a statistical fluctuation at $N_s = 12$ we can not yet say. Certainly, it would be interesting to fill in the point at 14 and repeat the calculation with a smaller lattice spacing where one can study more points along the curve. Of course, lattice calculations with boxes of size $< 1.8$ fm would not be sensitive to the effects near the minimum, and it is not clear how they could be used to make an extrapolation to the infinite volume limit.

In the Kogut-Susskind formalism, we have a second pair of local meson operators called the $\pi_2$ and $\rho_2$. (See Refs. 26, 27 and 28.) Going from $N_s = 8$ to 16, the $\pi_2$ mass decreases by 0.019(11), or 2.5%. This is a 1.7 $\sigma$ effect. We see a similar pattern for the $\rho_2$, where the mass decreases by 0.040(16), or 4.2%. The difference is 2.6 $\sigma$. The $\pi_2$ and $\rho_2$ masses as a function of lattice spacing are shown in Figs. 6 and 7.

## B. Flavor Symmetry of Pion and Rho

The $\pi_2$ and $\rho_2$ should be degenerate with the $\pi$ and $\rho$, respectively, if flavor symmetry is realized. At nearly this coupling, it is known from Ref. 10 that for the heavier mass there is considerable flavor symmetry breaking for the pion but not the rho. Of course, the errors were much larger in the previous work. In the current calculation, we find the $\rho_2$ is about 4% heavier than the $\rho$ for both quark masses with the largest lattice size. For every case, we find the $\rho_2$ is heavier than the $\rho$. For the pion, on the other hand, we find the $\pi_2$ to $\pi$ mass ratio increases as we decrease the quark mass. This is exactly what we expect since the $\pi$ is the Goldstone mode of the $U(1)$ symmetry of the Kogut-Susskind quarks, but the $\pi_2$ is not. The mass ratio is 1.69 and 2.16 for the heavier and lighter quarks, respectively. The HEMCGC collaboration also finds that this mass ratio is increasing as the quark mass decreases[29]. In contrast, the $MT_c$ collaboration[30] finds that $m_{\pi_2}/m_\pi$ is independent of quark mass, although it is not equal to one. The calculation here reaches a smaller value of $m_\pi/m_\rho$ than Refs. 29 and 30.

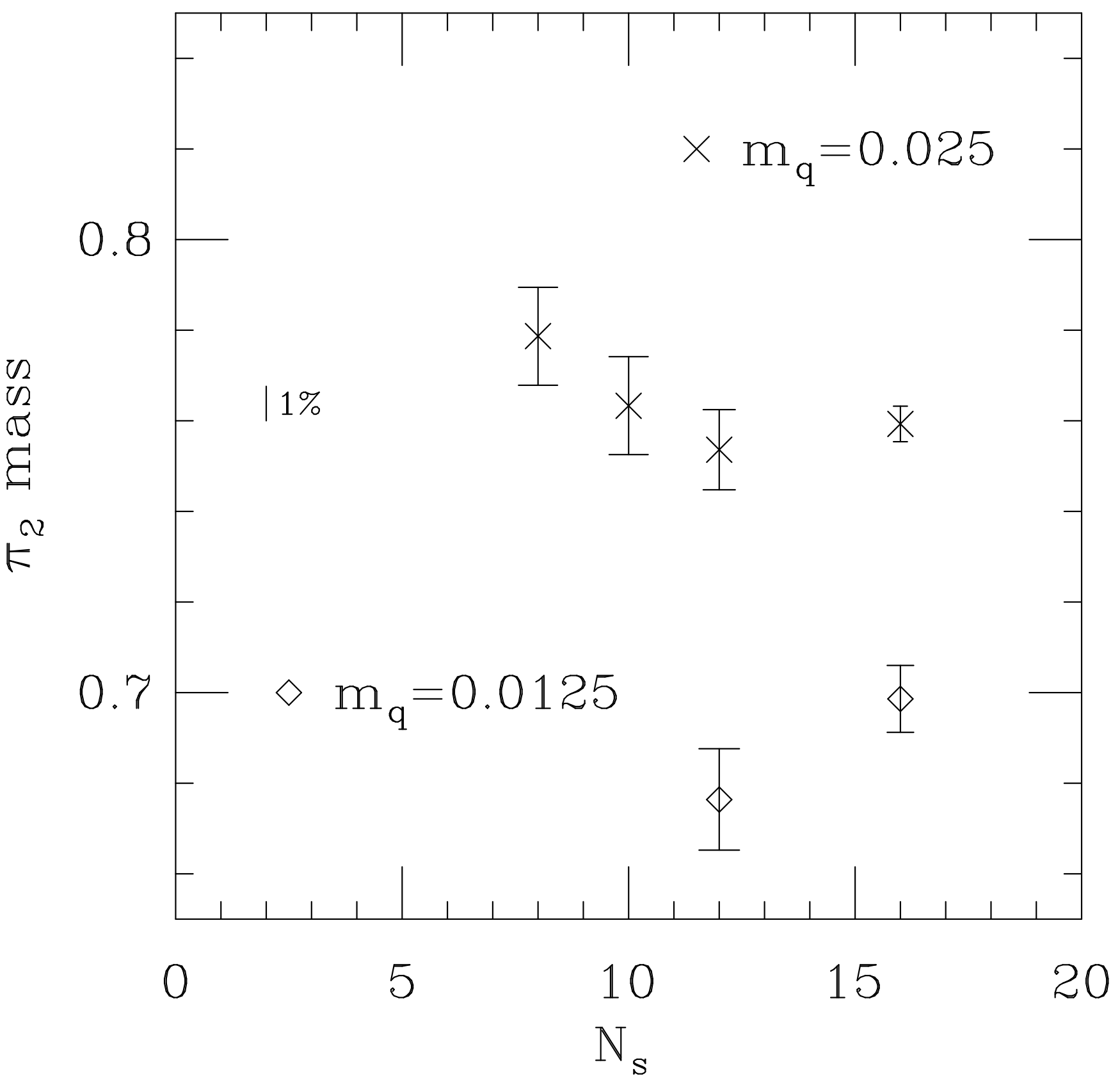

Fig. 6. Same as Fig. 3, but for the $\pi_2$.

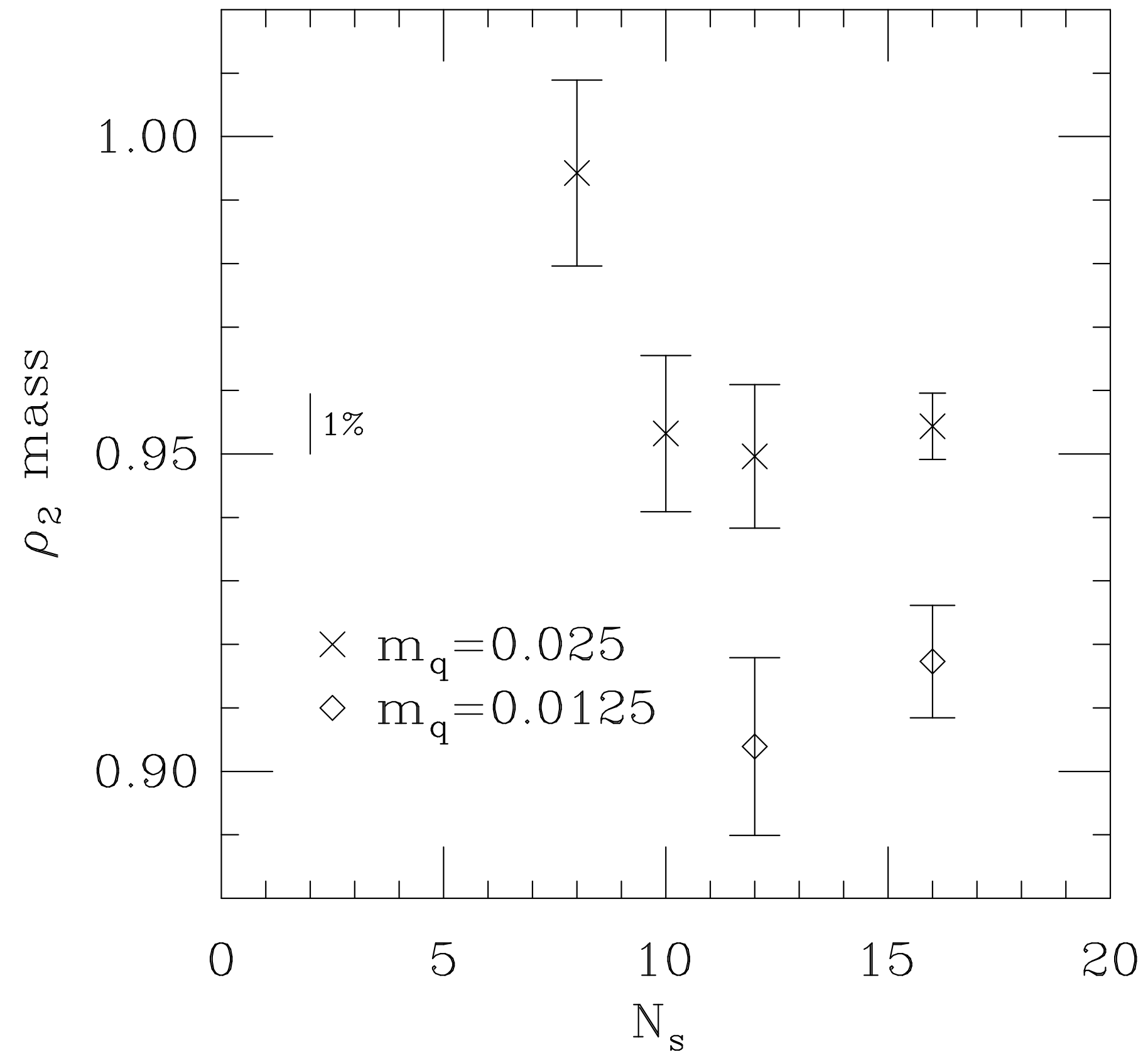

Fig. 7. Same as Fig. 3, but for the $\rho_2$.

C. Mesons with momentum

We have also measured masses for mesons with nonzero momentum. Our primary motivation was an attempt to see the effects of the decay $\rho \rightarrow 2\pi$, but the results can also be used to see how closely the energy-momentum relation for the lattice mesons approaches the continuum answer. In the real world, the rho decays strongly, with a width that is a large fraction of its mass. In lattice calculations with dynamical quarks, the pion mass has previously been large enough that the rho mass was below the two pion threshold. As the pions are made lighter, we expect the coupling of the rho and two pion states to affect the rho mass[14-17]. Indeed, when the two pion state is really lighter than the rho, the conventional lattice calculation of the rho mass would not find the rho at all, but rather a two pion state. Although it turns out that our attempt to see the effects of the $\rho$-$\pi$-$\pi$ coupling was unsuccessful, this issue must be sorted out before dynamical QCD simulations can produce realistic hadron phenomenology, and we hope that our attempt may be useful to further studies.

With Kogut-Susskind quarks, we expect to have four degenerate flavors of valence quarks in the continuum limit. Thus, we have sixteen $S$-wave pseudoscalar mesons (fifteen pions and an eta) and sixteen $S$-wave vector mesons. Each Dirac component of each of these quark flavors is a linear combination of the one component quark fields at each site of a $2^4$ hypercube on the lattice. The mesons are most easily described in the notation of Ref. 26, where they are created by the operator $\bar{\psi}\Gamma_{spin} \times \Gamma_{flavor}\psi$. Here $\Gamma_{spin}$ is a Dirac matrix giving the spin structure of the meson — $\gamma_5$ for the pion and $\gamma_i$ (or $\gamma_0\gamma_i$) for the rho. $\Gamma_{flavor}$ is a $4 \times 4$ matrix giving the flavor structure of the meson. With nonzero lattice spacing the pions are not degenerate, and only one pion, the $\gamma_5 \times \gamma_5$ pion, is an exact Goldstone boson as $m_q \to 0$. With the quark masses and lattice spacings we used, this member of the pion multiplet is considerably lighter than the other members and it is most promising to look at a vector meson which can decay into two Goldstone pions. As illustrated in Fig. 8, this requires that $\Gamma_{flavor}$ for the vector meson be $\mathbf{1}$, so that the trace over the flavor matrices does not vanish. Thus, we use the sink operator $\bar{\psi}\gamma_i \times \mathbf{1}\psi$, which is a nonlocal rho meson.

Two Harari-Rosner quark-flavor diagrams contribute to the decay of the rho meson, as shown in Fig. 9. The quark-antiquark flavor wave functions in the scalar and adjoint representation of SU(N) are the generators $T^a$ of U(N). They are normalized to $\mathrm{Tr}(T^a T^{a\dagger}) = 1$.

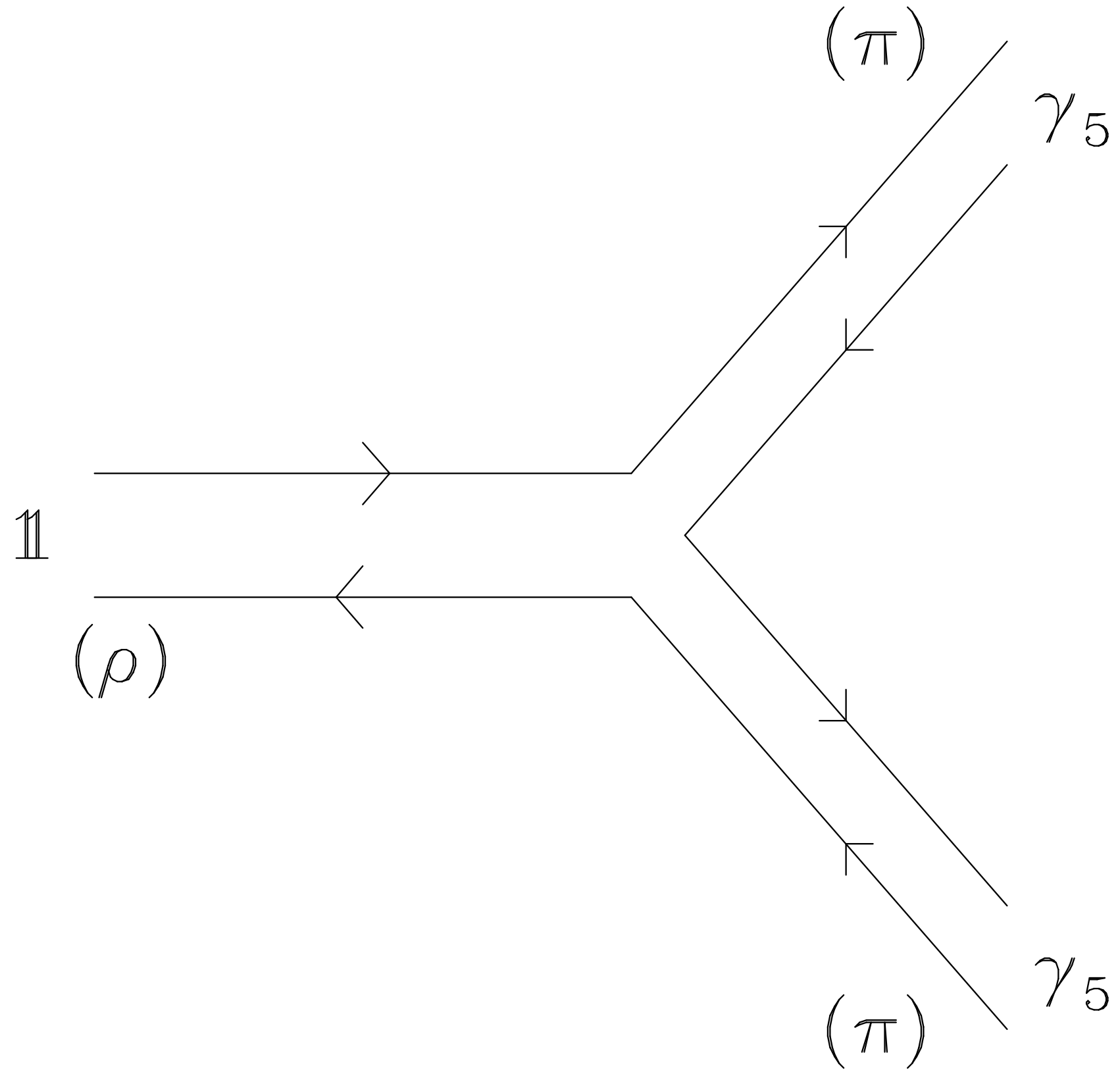

Fig. 8. Quark propagators for a rho decaying into two pions. For each particle the flavor matrix is shown. The Goldstone pions have a $\gamma_5$ in flavor space, requiring the decaying rho to have the unit matrix.

Then the flavor contribution to the partial width for a specific channel $a \to b\,c$ is

$$G^2_{a,bc} = g_V^2 |\mathrm{Tr}(T^a T^b T^c)|^2 - g_V^2 \mathrm{Tr}(T^a T^b T^c)\mathrm{Tr}(T^{b\dagger} T^{c\dagger} T^{a\dagger}), \tag{1}$$

the two terms coming from the two respective diagrams. For example, for the physical decay $\rho \to \pi\pi$ using SU(2) flavor, we have $G^2_{\rho,\pi\pi} = 2g_V^2$. With our conventions this is $g^2_{\rho\pi\pi}$ in the usual normalization. (See, for example, Ref. 31.) The experimental value is approximately $g^2_{\rho\pi\pi} = 36$.

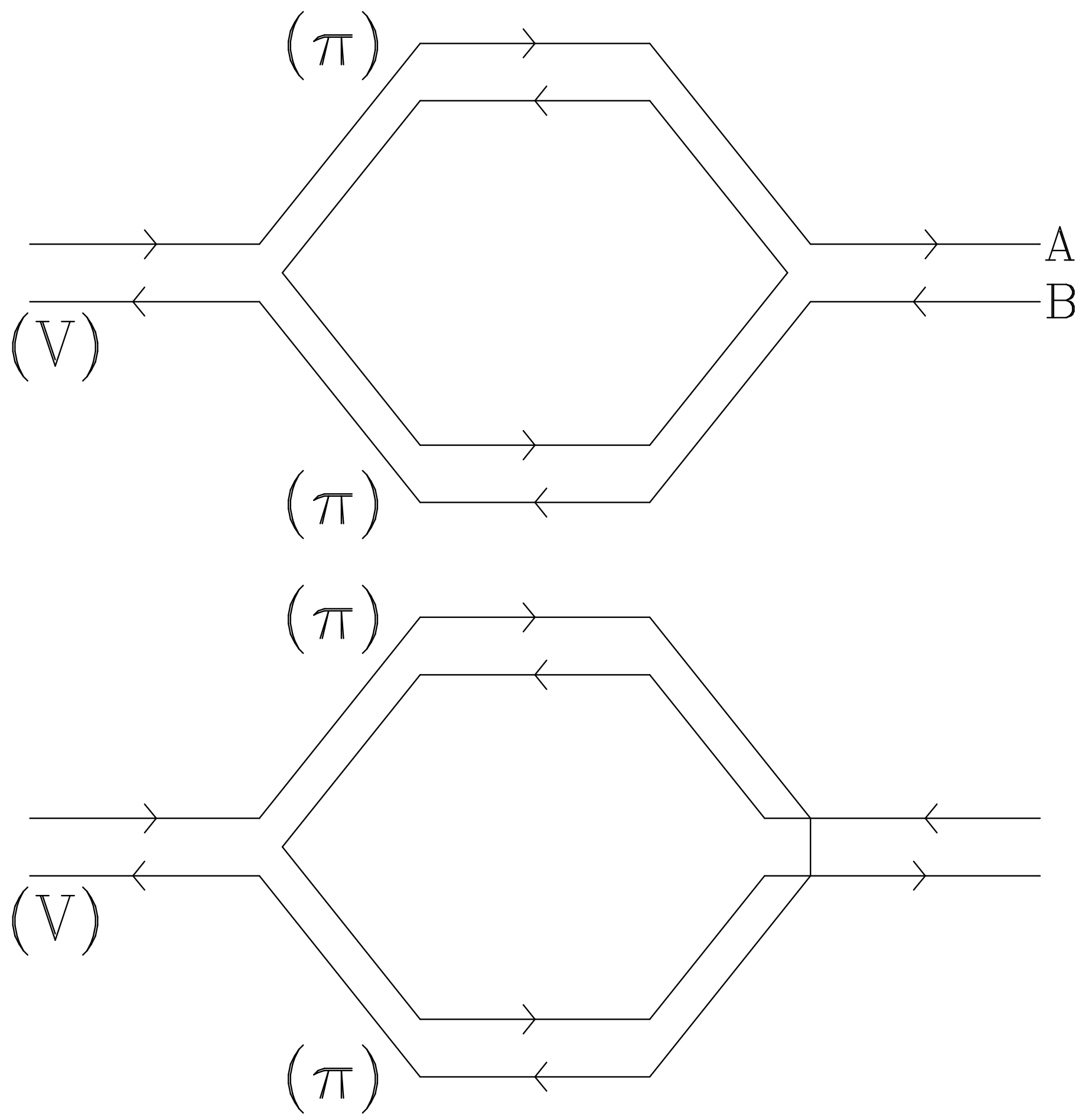

Fig. 9. Harari-Rosner graphs for the pion self energy contribution to the rho propagator. The second diagram can be obtained from the first by exchanging the two pions on one side of the diagram. For the meson propagators only the first diagram is computed here.

To make approximate contact with the staggered fermion simulation, which has an effective SU(4) flavor, we assume flavor independence and symmetry-breaking immunity. That is, the flavor part of the decay amplitude is the same, regardless of the number of flavors, and regardless of the degree of symmetry breaking apparent in the spectrum. Thus we keep the same value for $g_V^2$ and increase the number of flavors to 4, ignoring the effects of flavor symmetry-breaking mass shifts. We must now take account of three new circumstances. First, our vector meson propagator is composed of propagators of only one valence quark and one antiquark, corresponding to only the first Harari-Rosner diagram. Second, the virtual quark loop is suppressed by a factor of two because our two-flavor action includes the square root of the four-flavor fermion determinant. Finally, flavor symmetry breaking singles out a unique intermediate state involving the lightest pion channel, namely two $\gamma_5$ pions. Thus the flavor factor for the decay most strongly affecting the vector meson propagator in question is

$$G_{1,\gamma_5\gamma_5}^2 = g^2|\mathrm{Tr}(1\gamma_5\gamma_5)/8|^2/2 = g_V^2/8 \tag{2}$$

where the factor of 1/8 normalizes the three flavor wave functions and the factor of 1/2 comes from the square root of the fermion determinant. From the experimental rho width we then have $G_{1,\gamma_5\gamma_5}^2 \approx 36/16 \approx 2.3$.

Let us now turn to the kinematic factors. For this analysis we follow Ref. 16. The coupling of the rho to two pions is described by an effective Lagrangian $\rho_i\pi\partial_i\pi$ where $\rho_i$ is $\bar{\psi}\gamma_i\psi$. Converting this interaction to a lattice form, we calculate the dispersion relation for the vector meson with nonzero momentum $p_\rho$ along the $z$-axis. The dispersion relation up to one loop order in the pion-induced self energy is

$$G_L(\omega, p_\rho)^{-1} = 2(\cosh\omega - 1) - \mu_L^2 - 2(1 - \cos p_\rho) - G_{a,bc}^2 S_L(\omega, p_\rho) = 0 \tag{3}$$

where the subtracted self energy $S_L(\omega, p_\rho)$ is given in terms of the unsubtracted self energy $\Pi_L(\omega, p_\rho)$ by

$$S_L(\omega, p_\rho) = \Pi_L(\omega, p_\rho) - \Pi_L(0, p_\rho) - \frac{\partial^2}{\partial\omega^2}\Pi_L(0, p_\rho)(\cosh\omega - 1). \tag{4}$$

The unsubtracted self energy is

$$\Pi_L(\omega, p_\rho) = -\frac{1}{L^3} \sum_{k_x,k_y,k_z} \left\{ \frac{P}{4 \sinh E_1 [\cosh E_2 - \cosh(\omega - E_1)]} \right. \\ \left. + \frac{P}{4 \sinh E_2 [\cosh E_1 - \cosh(\omega + E_2)]} \right\} \tag{5}$$

with $(k_x, k_y, k_z) = 2\pi(n_x, n_y, n_z)/L$ for integer $n_x, n_y, n_z \in [0, L-1]$ and polarization factor

$$P = [(\sinh E_1 - \sinh E_2)\epsilon_t + (\sin k_{1x} - \sin k_{2x})\epsilon_x + (\sin k_{1y} - \sin k_{2y})\epsilon_y + (\sin k_{1z} - \sin k_{2z})\epsilon_z]^2. \tag{6}$$

The momenta of the intermediate pions are given by $\mathbf{k}_1 = \mathbf{k}$ and $\mathbf{k}_2 = \mathbf{p}_\rho - \mathbf{k}$ and the energies by

$$\cosh E_1 = 1 + \mu_\pi^2/2 + (1 - \cos k_x) + (1 - \cos k_y) + (1 - \cos k_z) \tag{7}$$

and

$$\cosh E_2 = 1 + \mu_\pi^2/2 + (1 - \cos k_x) + (1 - \cos k_y) + [1 - \cos(p_\rho - k_z)]. \tag{8}$$

Finally, for polarization parallel to the momentum, the polarization vector for the initial rho meson is given by

$$\epsilon_{||t} = -\sin p_\rho / \sqrt{\sinh \omega^2 + \sin p_\rho^2} \tag{9}$$

and

$$\epsilon_{||z} = \sinh \omega / \sqrt{\sinh \omega^2 + \sin p_\rho^2} \tag{10}$$

with $\epsilon_{||x} = \epsilon_{||y} = 0$, and for polarization perpendicular to the momentum by

$$\epsilon_{\perp x} = 1 \tag{11}$$

and $\epsilon_{\perp t} = \epsilon_{\perp y} = \epsilon_{\perp z} = 0$.

As seen from the expression for the polarization factor $P$ above, the amplitude for the $\rho$-$\pi$-$\pi$ coupling vanishes as the relative momentum of the pions vanishes. Thus, at least one of the pions must have nonzero momentum. If we consider only momenta of zero or $2\pi/L$, the most favorable case for the rho mass to exceed or equal the two pion threshold is a rho with momentum $2\pi/L$ mixing with two pions with momenta 0 and $2\pi/L$. In order to get a good overlap of our wall source with a rho with nonzero momentum, we use a different wall source for the quark and antiquark propagator. For the quark propagator, we use our usual corner wall source. (We use the lattice Coulomb gauge.) Roughly speaking, this produces a quark with momentum zero. For the antiquark source, we set the $(0,0,0)$ elements of the cubes to $\cos(2\pi x/L)+\cos(2\pi y/L)+\cos(2\pi z/L)$, to produce an antiquark with one unit of momentum.

Consider a rho with momentum in the $z$ direction, $\vec{p} = \frac{2\pi}{L}\hat{z}$. This mixes with a two pion state where the pions have momenta 0 and $\frac{2\pi}{L}\hat{z}$. Again from the polarization factor $P$, we see that the "parallel" rho $\bar{\psi}\gamma_z\psi$ should couple to this two pion state while the "perpendicular" rhos, $\bar{\psi}\gamma_x\psi$ and $\bar{\psi}\gamma_y\psi$ should not. The signal that we look for is a difference in the mass of the parallel and perpendicular rho mesons (averaged over momenta in the $x$, $y$ and $z$ directions).

The nonlocal rho we are using involves the product of quark and antiquark propagators displaced by one lattice link. In taking these products, we parallel transport along the lattice link. As a control on our calculation, we also calculated the propagator for a nonlocal pion involving propagators displaced by one lattice link, $\bar{\psi}\gamma_{5(spin)}\times\gamma_5\gamma_{i(flavor)}\psi$. When this pion has nonzero momentum, we may also separate it into "parallel" and "perpendicular" components depending on whether $i$ is the direction of the momentum or perpendicular to the momentum. We do not expect to see differences between these propagators.

Table III contains results for the local and one-link rho and pion with zero and nonzero momenta. These propagators were obtained from 200 stored lattices at $am_q = 0.025$, with four consecutive lattices blocked together. For $am_q = 0.0125$, all 611 lattices were used with a block size of eight. In both cases, we used four wall sources per lattice. We chose different fits for the local pion and rho in Table III than Table II because for this comparison we wanted to use the same distance range for both zero and nonzero momentum.

For $am_q = 0.025$, the nonzero momentum $\gamma_i \times \mathbf{1}$ rho is very close to the threshold for decay into two Goldstone pions: $0.448 + 0.584 = 1.03$, while for $am_q = 0.0125$ this rho is above the two pion threshold: $0.324 + 0.501 = 0.83$. Evidently, Table III shows no significant difference between the parallel and perpendicular $\gamma_i \times \mathbf{1}$ rho masses. It is interesting to observe that all of the above masses come reasonably close to the continuum dispersion relation $m^2_{p=2\pi/L} = m^2_{p=0} + (2\pi/L)^2$.

Is this result consistent with what is expected from the experimental value of the coupling? Our assumptions of flavor independence and symmetry-breaking immunity fix the mixing strength $G^2$ between the bare rho and pion channels in terms of the experimental width of the rho. We make the final assumption that our source and sink couple more strongly to the bare rho meson than to any of the bare two pion channels. This assumption is borne out in our conclusions that mixing to the pion channels is inherently weak. Then the lattice propagator for the observed vector meson channel is simply proportional to $G_L(\omega, p_\rho)$. This propagator has a series of poles corresponding to the renormalized rho meson and the several pion channels. The $n$th pole in $G_L(\omega, p_\rho)$ at $\omega = m_n$ and residue $\beta_n$ contributes a term

$$N\beta_n\{\exp(-m_n t) + \exp[-m_n(L-t)]\} \tag{12}$$

to the time dependence of the correlator. Thus with our final assumption, the relative strengths of all spectral components can be predicted from the ratios of the residues in $G_L(\omega, p_\rho)$. Shown in Table IV are results of a calculation of pole positions and residues for parameters appropriate to the two quark masses considered and with illustrative choices for the value of the bare rho mass. In the case $\mu_L^2 = 1.0$ and $\mu_\pi = 0.4478$, appropriate to $am_q = 0.025$, in the upper part of the table, the lowest bare two pion pole in the parallel channel (1.0297) is very close to the bare rho mass (1.0366) resulting in strong mixing. Nonetheless, the splitting of the two resulting states (1.0228 and 1.0349) is far too small to be distinguishable in our simulation. The other two-pion states are so weakly coupled that they have negligible effect on the correlator. Thus the whole spectrum would appear to our analysis as a single state with mass approximately 1.036. In the corresponding perpendicular channel, the renormalized rho is the strongest state and dominates the correlation with its mass of 1.026. Already without attempting to further tune the bare rho mass, we see that our $am_q = 0.025$ results are consistent with this scenario. Turning next to the case $\mu_L^2 = 0.96$ and $\mu_\pi = 0.3242$ appropriate to $am_q = 0.0125$ in the lower part of the table, we note that the lowest bare two pion state is well below the bare rho. However, it is so weakly mixed that it would be invisible to our analysis. We can only say that our null results are consistent with expectation.

Thus a variety of circumstances conspire to make it extremely difficult to observe the effects of rho decay on the rho propagator. First, flavor symmetry breaking inherent in the staggered fermion scheme reduces the number of available channels by nearly a factor of 15. Second, with our lattice dimensions, finite size effects produce such a coarse spacing of the discretized continuum that too few $\pi - \pi$ states are available for mixing. The result is that when the bare rho meson is close enough to mix strongly with a $\pi - \pi$ state, the consequent splitting is too small to resolve, and when it is farther away, the mixing is too weak to produce a signal. In either case, the resulting spectral component is practically indistinguishable from the unmixed rho meson.

These results underscore important limitations inherent in the staggered fermion scheme with present lattice sizes. To make further progress obviously requires larger lattices and much weaker coupling to give a more realistic representation of the two pion continuum. In the mean time, more direct methods observing an explicit $\rho \to \pi\pi$ transition hold more promise for making a comparison with experiment[32].

## D. Another Source and the Delta Baryon

In addition to the "corner wall" source used for most of our mass estimates, we have also used "even" and "odd" wall sources[33,34], which couple to the local nucleon, a nonlocal nucleon and the delta. The even wall source is constant over all sites on one time slice, while the odd wall source assigns $+1$ to even spatial sites and $-1$ to odd sites (in Coulomb gauge). The product of three even sources contains representations of the discrete lattice symmetry group sufficient to yield the desired states[33,35]. We have calculated propagators from this source on 200 stored $16^3 \times 24$ lattices with $am_q = 0.025$, and on the entire data set with $am_q = 0.0125$. Simulations with two flavors of Kogut-Susskind quarks at $6/g^2 = 5.6$ show that the effective masses of the nucleons from these two sources differ out to fairly large separation[34].

In Fig. 10, we show the effective masses for the local nucleon from the corner wall source and both the local and nonlocal nucleons from the even wall source for $am_q = 0.0125$. Results were similar at $am_q = 0.025$. We see that at short distances the nucleons from the even source have smaller effective masses than the nucleons from the corner source, as found by the HEMCGC group[34]. However, by the time the effective mass has leveled off the two sources give the same mass within statistical errors.

The even source couples to the delta. For $am_q = 0.025$ reasonable fits are obtained. We may estimate the $\Delta$ mass as 1.43(4) from a two-particle fit to distance range 2 to 9, which has $\chi^2 = 3.9$ for four degrees of freedom(CL=0.42). For $am_q = 0.0125$, we did not get good fits for the $\Delta$ and we are unable to quote a mass.

## E. The Edinburgh Plot

In Fig. 11, we display our Edinburgh plot for the current runs. We note that to determine the error in the nucleon to rho mass ratio, we have just added the errors as if there is no correlation between the two masses. When we looked at correlations between the two masses for $N_s = 16$, $am_q = 0.0125$, we found the the correlation varied between $-0.06$ and 0.08 depending on the block size used in averaging the data, so there is no significant effect on the error of the ratio. Looking first at the four points for $am_q = 0.025$, we see that there is some finite size effect between the smallest size $N_s = 8$ and the others. However, given the size of the error bar for $N_s = 8$, this is only about a one standard deviation effect. The octagon, corresponding to $N_s = 16$, is our best estimate of the ratio for this quark mass. For the lighter mass, 0.0125, we have the two points to the left with $m_\pi/m_\rho \approx 0.36$. The fancy cross, corresponding to $N_s = 12$ is, unfortunately, the lower of the two values. The reader will recall that for the lighter quark mass the rho mass fell as $N_s$ was increased from 12 to 16, unlike the heavier quark mass. We had hoped that the nucleon to rho mass ratio would drop significantly below 1.5 as we decrease the quark mass. Of course, we are working at a fairly strong coupling and don't expect to get the physical ratio (shown as the plus sign at the lower left) in the chiral limit. A recent quenched calculation with Wilson quarks has investigated how the ratio decreases with the lattice spacing[6]. After extrapolation of hadron masses to the chiral limit and taking account of finite volume effects, this work finds that the nucleon to rho mass ratio approaches the physical value when extrapolated to zero lattice spacing. With current computer resources, it is possible to work at smaller lattice spacing within the quenched approximation. To study the chiral limit, we have fit the rho and nucleon masses and $m_\pi^2$ to linear functions. In Fig. 12, we show the mass ratios from the resulting fits with $N_s = 12$ and 16 for the lighter quark mass. We see that the extrapolations differ by about two standard deviations, and the extrapolation based on $N_s = 16$ is not significantly different from 1.5. Figure 13 compares the current mass ratios with those based on the HEMCGC calculations [29] at $6/g^2 = 5.6$, and others done at 5.7 by the Columbia[37] and Kyoto-Ibaraki-Kofu [36] groups. The figure details which points correspond to which simulations. This figure shows the chiral extrapolation for the $6/g^2 = 5.7$ results of Ref. 36. Here, $m_N/m_\rho$ falls to about 1.43, which is smaller than our stronger coupling result. Nevertheless, this is still far from the physical ratio of 1.22.

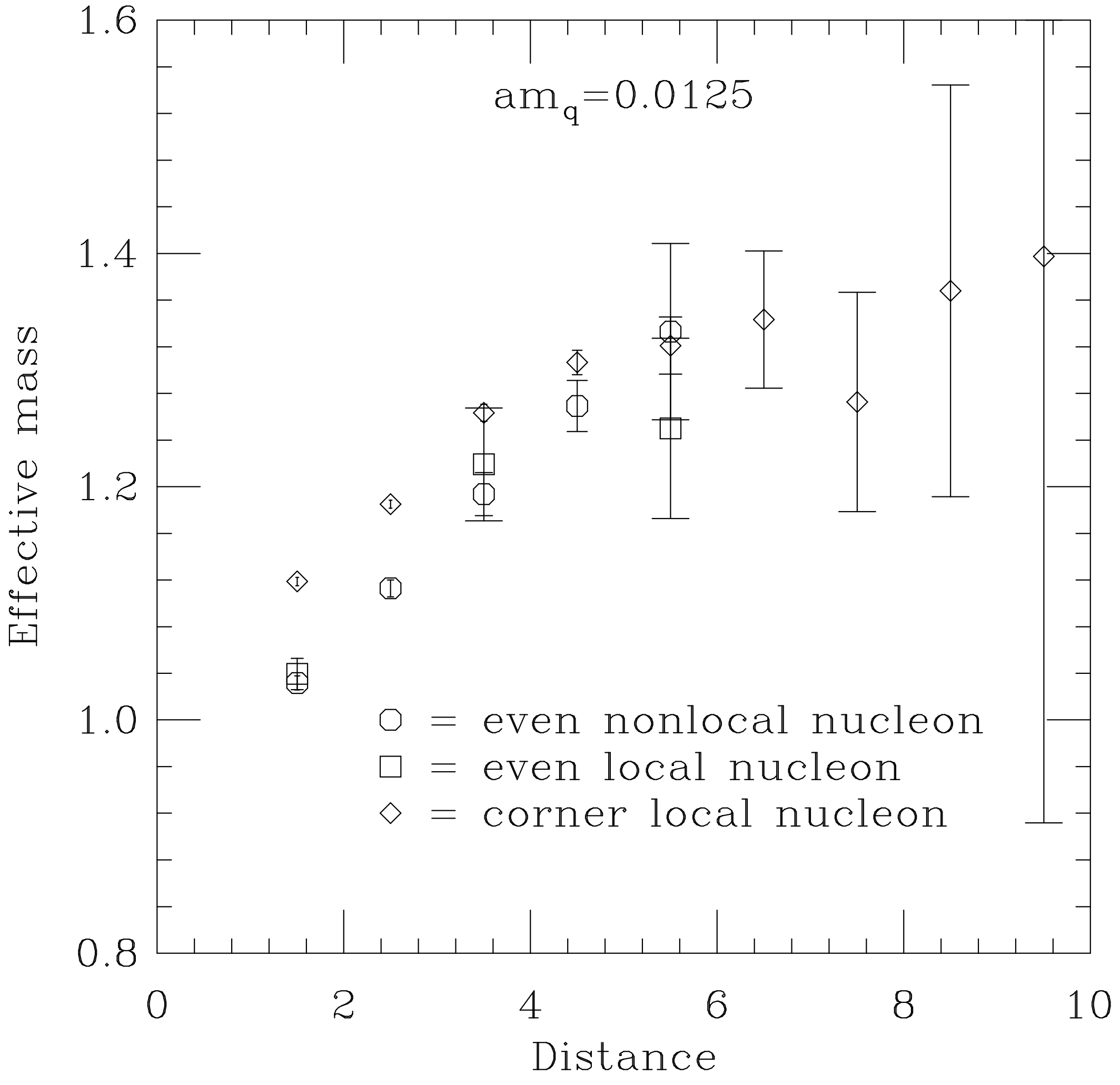

Fig. 10. The effective nucleon mass as a function of the distance from the source plane for $am_q = 0.0125$. Close to the source, the corner wall source and even wall source given different values; however, the two sources agree by the time the effective masses have leveled off.

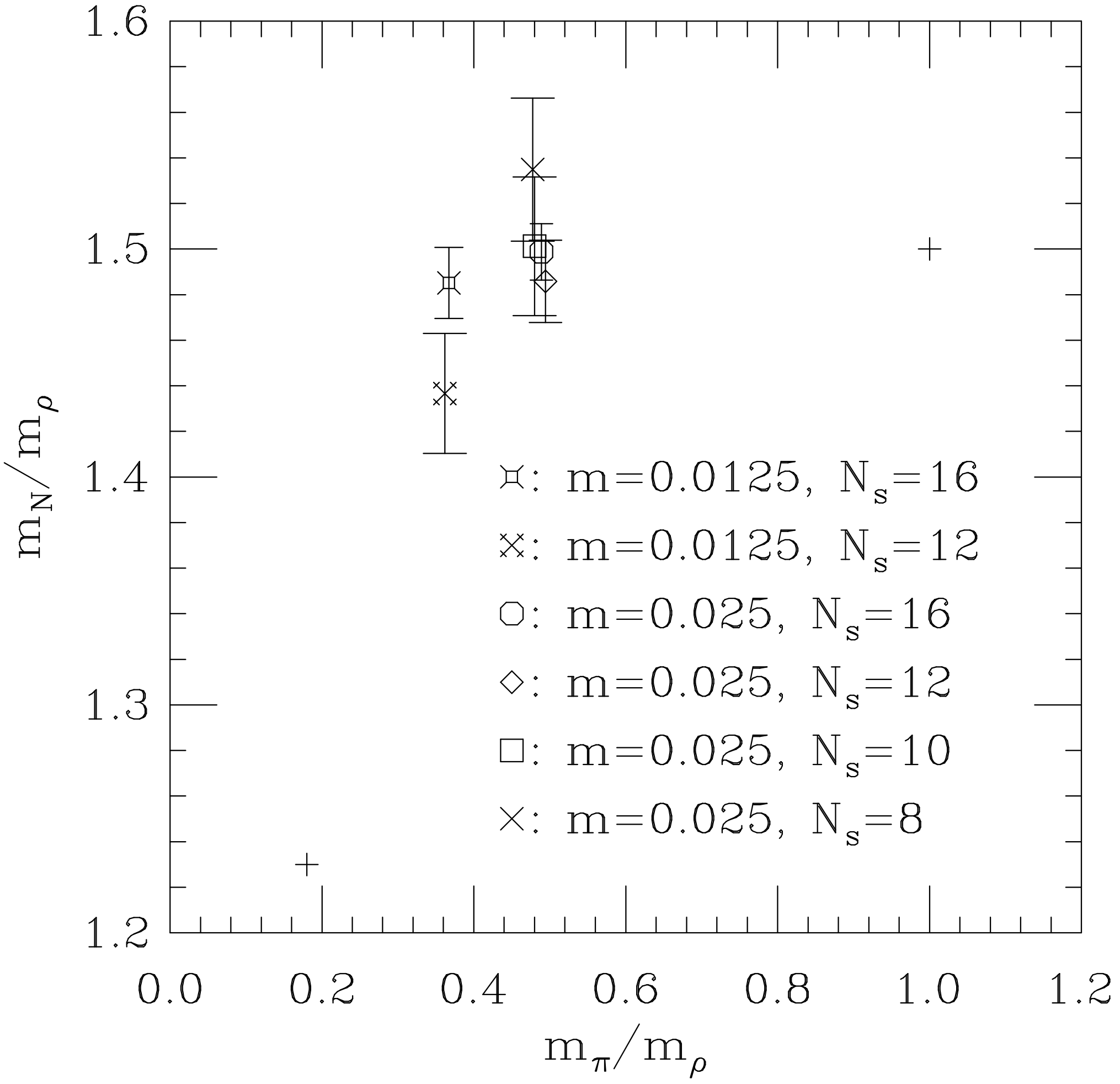

Fig. 11. Edinburgh plot showing all six mass and volume combinations considered here.

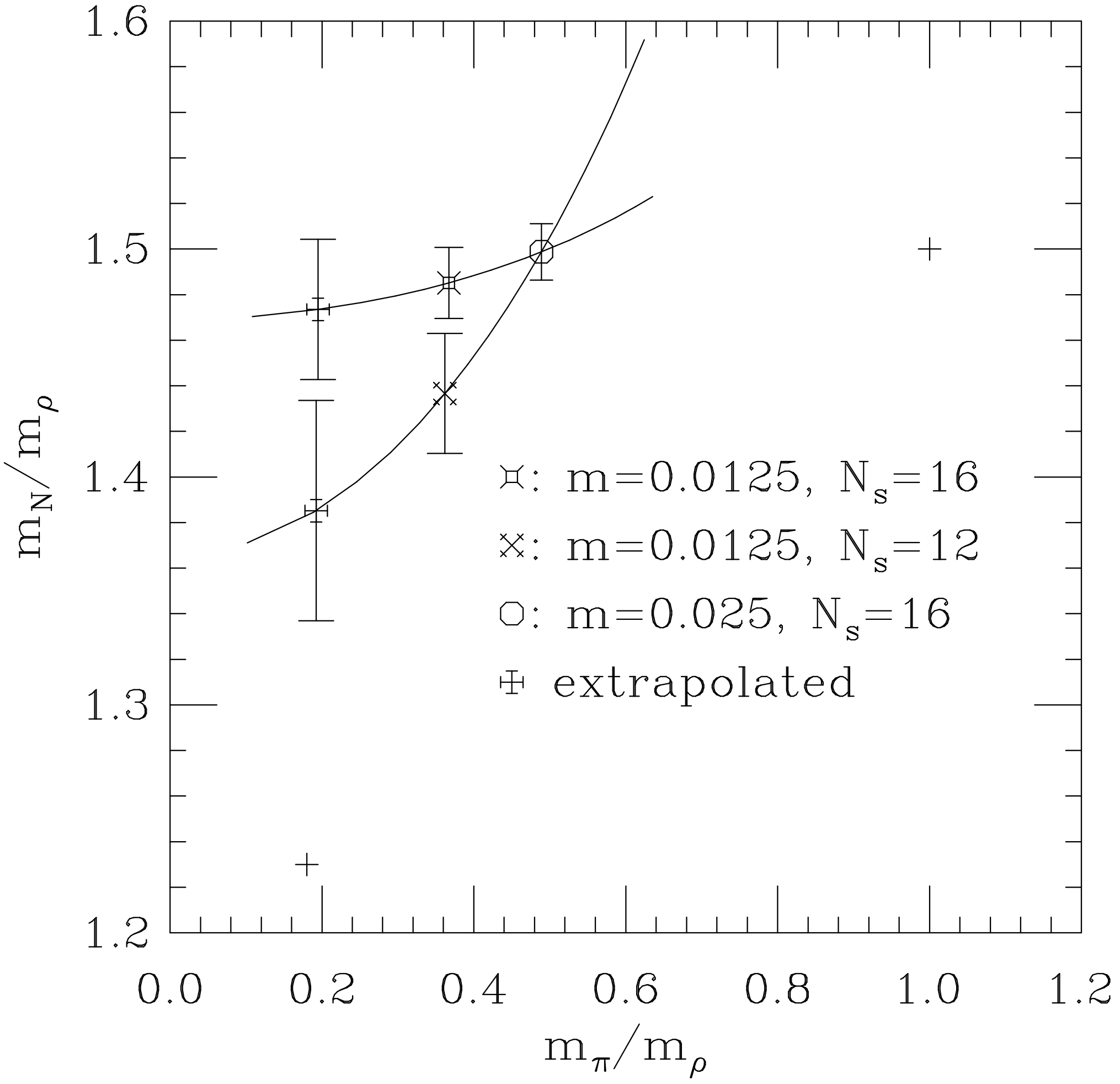

Fig. 12. Edinburgh plot showing extrapolations to lighter quark mass. The $N_s =$ 16 results for the heavier quark mass and either size for the lighter quark mass are used as a basis for the extrapolations.

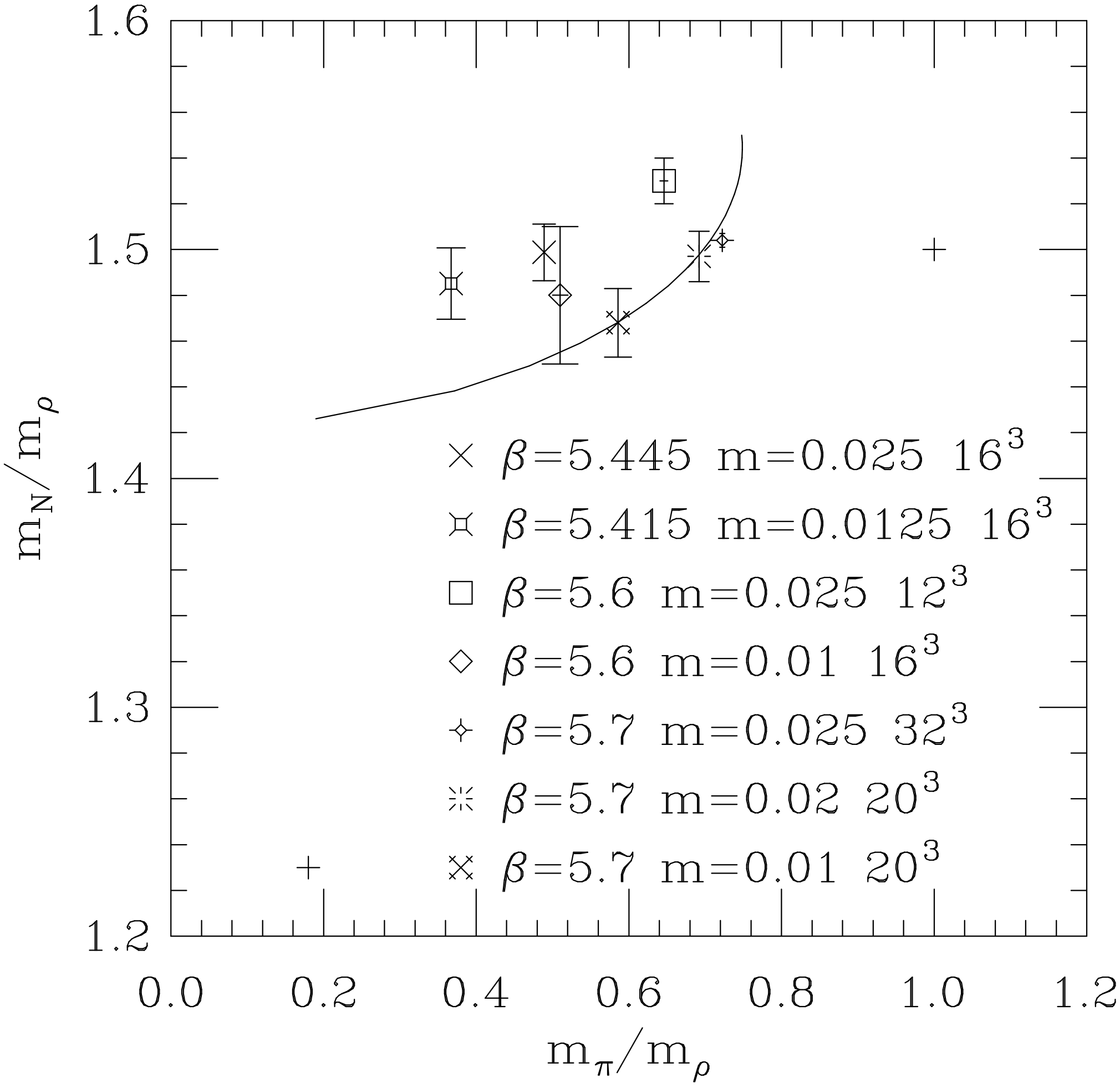

Fig. 13. Edinburgh plot comparing our results with those of previous spectrum calculations with $6/g^2 = 5.6$ and 5.7. The line is the chiral extrapolation of the results of Ref. 36.

## IV. HADRON MASS DERIVATIVES

Hadron spectrum calculations require enormous computer resources, so it is important to maximize the information extracted. Techniques exist for using a simulation with one coupling to infer information about nearby couplings[38]. Since the current lattice simulation uses an unphysically large quark mass, it is of interest to attempt an extrapolation of our measurements to small quark mass. In particular, we would like to see hadron mass ratios approach their physical values as the quark mass is lowered. This might be accomplished by computing hadron mass derivatives. These are determined by taking numerical differences of masses obtained from fits to propagators with slightly different quark masses. One of the propagators is the usual operator measured on the lattice. The second is formally obtained from the first by a Taylor series expansion:

$$\langle G(m+dm)\rangle \approx \langle G(m)\rangle + \frac{\partial \langle G(m)\rangle}{\partial m} dm \tag{13}$$

where $m$ is the quark mass. The mass dependence of the propagator is given explicitly by

$$\langle G(m)\rangle = \frac{\int [dU] G(m) e^{-S(m)}}{\int [dU] e^{-S(m)}} \tag{14}$$

where the integrals are over the gauge field $U$ and $S = S_{gauge} - \frac{n_f}{4}\mathrm{Tr}\left(\log M(m)\right)$ is the QCD action. Here $M$ is the usual fermionic matrix with diagonal elements $am$, $n_f$ is the number of quark flavors, and $V$ is the space-time volume of the lattice. When measuring operators on the lattice, we are free to choose the mass appearing in the operator differently from the mass used in the simulation. The first is the valence quark mass, $m_v$, and the second is the dynamical quark mass, $m_d$. This distinction is, of course, not physical. We imagine that the propagating quarks with mass $m_v$ that make up our hadrons are moving in a background field generated by gluons and dynamical quarks with mass $m_d$. By taking $m_d$ and $m_v$ to be independent, we get two contributions to the hadron mass derivative, $\partial m_h/\partial m_d$ and $\partial m_h/\partial m_v$. To compute $\partial \langle G\rangle / \partial m_v$, $\langle G(m+dm_v)\rangle$ is measured directly on the original lattice ($m_v = m_d = 0.0125$) with $dm_v = 0.01 m_d$. We checked to see that this change was in the linear region. Because we use the same set of lattices for both valence quark masses, the small changes in the hadron masses are not overwhelmed by statistical fluctuations. To find the propagator with slightly different $m_d$, we must compute $\partial \langle G(m) \rangle / \partial m_d$ which turns out to be a simple correlation of $G$ with $\bar{\psi}\psi$:

$$\frac{1}{V}\frac{\partial \langle G(m) \rangle}{\partial m_d} = \frac{1}{V}\frac{\partial}{\partial m_d}\Big(\frac{\int [dU] G(m_v) e^{-S(m_d)}}{\int [dU] e^{-S(m_d)}}\Big) \tag{15}$$

$$= \langle G\bar{\psi}\psi \rangle - \langle G \rangle \langle \bar{\psi}\psi \rangle \tag{16}$$

with

$$\bar{\psi}\psi = \frac{n_f}{4V} \mathrm{Tr}\, M^{-1}. \tag{17}$$

We use a Gaussian random estimator to compute $\mathrm{Tr}\, M^{-1}$. The trace of any matrix $A$ can be computed by introducing the Gaussian integral

$$\mathrm{Tr}\, A = \frac{\int \prod_i dR_i\, dR_i^* \;\; R_j^* A_{jk} R_k \exp(-R^* \cdot R)}{\int \prod_i dR_i\, dR_i^* \; \exp(-R^* \cdot R)}. \tag{18}$$

Inserting the above into $\langle \bar{\psi}\psi \rangle$, and dropping the indices that run over position, spin and color, we find

$$\langle \bar{\psi}\psi \rangle = \frac{n_f}{4V} \frac{\int [dU] \int [dR\, dR^*]\, (R^* M^{-1} R)\, e^{-R^* \cdot R}\, e^{-S} / \int [dR\, dR^*]\, e^{-R^* \cdot R}}{\int [dU] e^{-S}} \tag{19}$$

Instead of actually integrating over $R$ and $R^*$, one or more random vectors with probability distribution function $\exp(-R^* \cdot R)$ are created and $R^* M^{-1} R$ is calculated for each random vector. The estimate of the trace is the average over the vectors. Our measurements described below were made using 12 random vectors for $\bar{\psi}\psi$ on each lattice. The $\pi$, $\rho$, and $N$ propagators were computed using six corner wall sources per lattice. Unless stated otherwise, the results given below were computed from correlated fits to two particle propagators (of opposite parity for the $\rho$ and $N$) where the data were grouped in blocks of six propagators per lattice times eight lattices (48 in all). All measurements were made on the $N_s = 16$ lattices with the lighter quark mass, $am_q = 0.0125$.

The various mass derivatives measured this way are given in Table VI. The errors are from an ordinary jackknife estimate where the data were grouped in blocks of 1/5 of the number of lattices and one block was omitted from each fit. The errors were insensitive to block size. $\chi^2$ and CL refer to fits to the original propagator with no blocks omitted.

The valence mass derivatives are quite good. The values shown are from the best fits over a wide distance range. Values obtained over different ranges with one or two particle fits agreed within errors to the values shown. The $\pi$ is seen to have the largest derivative which is reasonable since it is expected to go like $\sqrt{m_q}$ as $m_q$ goes to zero. Ratios of $N$ and $\rho$ and $\rho_2$ derivatives fall between 1.36 and 1.86 for the various ranges given in Table VI, which is in agreement with the naive expectation of 1.5 based on a simple constituent mass model of the hadrons.

The dynamical mass derivatives obtained from the simulation are less reliable. While the derivatives agree within errors for different fit ranges (except for the $\rho_2$), there is a large jump as the minimum distance increases. In fact, the derivatives double over the fit ranges shown. As a check, the correlations were recomputed using $\bar{\psi}\psi$ only over the fit range instead of the entire lattice. This produced similar results. To investigate further, we computed the correlation of the effective $\pi$ mass with $\bar{\psi}\psi$ at each time slice. The result is shown in Fig. 14a. The effective mass is positively correlated with $\bar{\psi}\psi$ on one side and anticorrelated on the other. This is consistent with Fig. 14b, which shows that the pion propagator for a given time slice is positively correlated with $\bar{\psi}\psi$ on both sides of the time slice, but the slope of the correlation with $t$ changes at the time slice.

With the derivatives in hand, we can attempt an extrapolation to small quark mass. Our values give no indication that at this lattice spacing the physical values of the hadron masses can be reached by simply lowering the quark mass. From Table VII, we see that our best value of $\frac{\partial}{\partial m_q}(\frac{M_N}{M_\rho}) = (\frac{\partial}{\partial m_v} + \frac{\partial}{\partial m_d})\frac{M_N}{M_\rho}$ is -2.0(5.5) and our best value for $\frac{\partial}{\partial m_q}(\frac{M_N}{M_{\rho_2}})$ is -9.9(4.6) where the errors are jackknife estimates. To reach the physical value $M_N/M_\rho =$

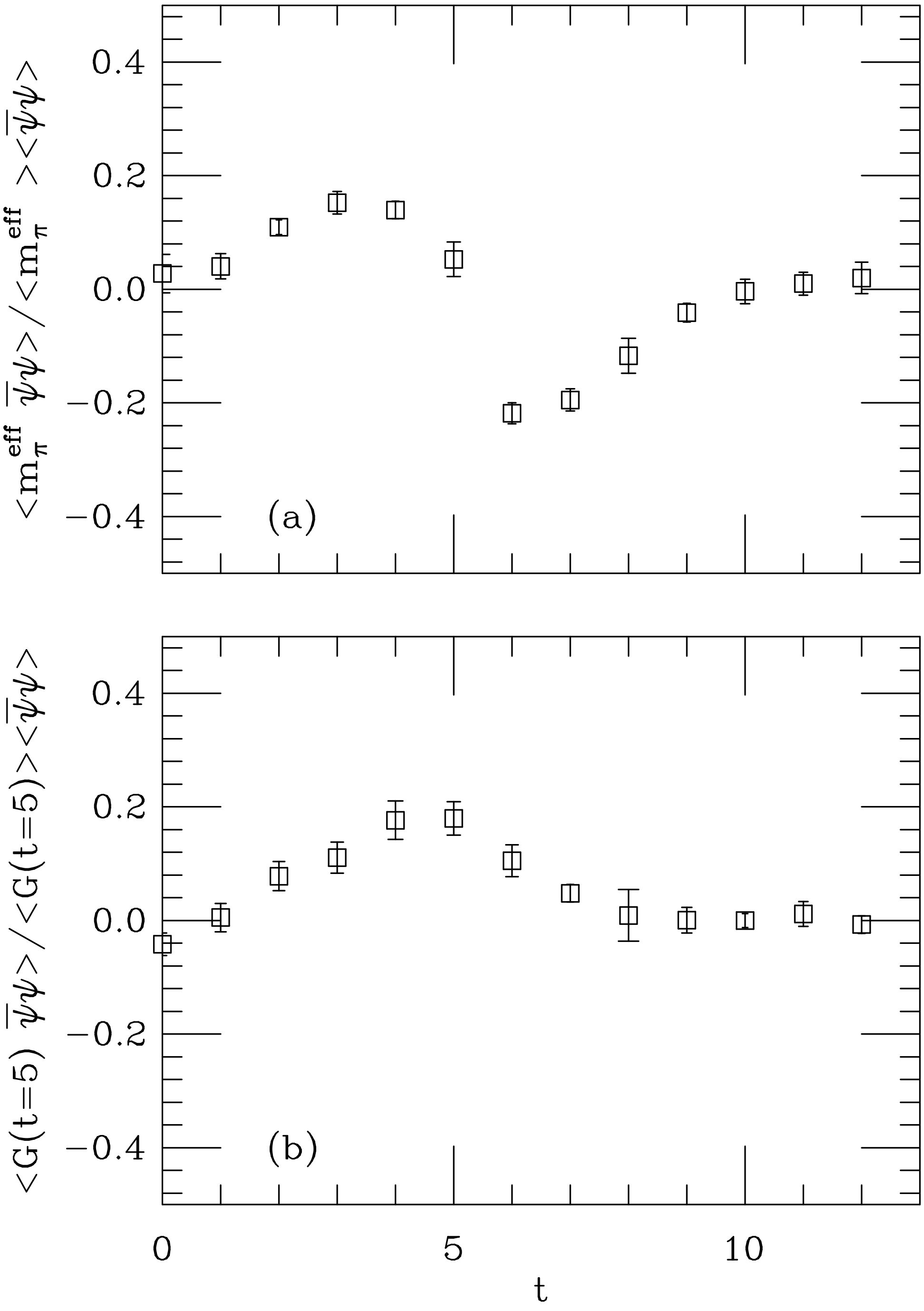

Fig. 14. (a). Correlation of the $\pi$ effective mass with $\bar{\psi}\psi(t)$. The effective mass was evaluated at distance of 5 time slices from the source. Error bars are jackknife estimates. (b). Correlation of the pion propagator at distance 5 with $\bar{\psi}\psi$ on different time slices.

1.23 at zero quark mass a derivative of approximately $(1.485 - 1.23)/0.0125 = 20.4$ is required.

The nucleon mass derivative is also related to the pion-nucleon sigma term, $\sigma_{\pi N}$, which has been previously computed on the lattice with dynamical quarks[39,40]. With $\sigma_{\pi N} = m\partial m_N/\partial m$, we find $\sigma_{\pi N}/m = 17.5(4.3)$, $24.7(10.6)$ for fitting ranges 3–9 and 4–11, respectively. The ratio of contributions to $\sigma_{\pi N}$ from dynamical and valence quarks is given by the ratio $\sigma_{\pi N}/\sigma_{\pi N}^{val}$ which we find to be 1.9(4) and 2.8(2.3) for the two fitting ranges mentioned. These values were computed at $\beta = 5.415$ and $am = 0.0125$. Our results are in rough agreement with those of Patel, who finds that $\sigma_{\pi N}/\sigma_{\pi N}^{val}$ is between 2 and 3 with dynamical Wilson quarks. The values given here are for $m_\pi^2/m_\rho^2 = 0.136$, and taken with those from Patel, show that the dynamical and valence quark mass contributions to $\sigma_{\pi N}$ seem to be independent (within errors) of $m_\pi^2/m_\rho^2$ over a wide range (0.136–0.8).

## V. COMPARISON OF WILSON AND KOGUT-SUSSKIND EDINBURGH PLOTS

An issue of crucial importance to lattice simulations is whether there are differences between Wilson and Kogut-Susskind quarks. In the quenched sector, it is well known[41] that there is a scale difference in the hadron masses that decreases as $6/g^2$ is changed from 5.7 to 6.0. Values of $m_N/m_\rho$ greater than 1.5 have been seen for Wilson quarks when $m_\pi/m_\rho \geq 0.75$; however, the evidence for this peak in the Edinburgh plot is much weaker for Kogut-Susskind quarks. This is partially a matter of there being less work with Kogut-Susskind quarks for these values of $m_\pi/m_\rho$. With dynamical quarks especially, the expectation has been that the interesting region is where $m_\pi/m_\rho$ is small. Since the determination of the nucleon to rho mass ratio at the chiral limit involves an extrapolation that may involve rather large quark masses, it may be important to compare the two fermion regularizations for all values of $m_\pi/m_\rho$. To this end, we have calculated the hadron spectrum with various large valence quark masses (hopping parameters) for both regularizations in our ensemble of gauge fields generated with dynamical quarks. For Kogut-Susskind quarks, our valence quark masses ranged from 0.2–1.6, in increments of 0.2 and we show plots for the $\rho$ and $\rho_2$. For the Wilson valence quarks, we used $\kappa$ values of 0.14, 0.15, 0.155, 0.16, and 0.165. The results are shown in Fig. 15.

For the $\rho_2$ (octagons), as the quark mass increases, $m_N/m_{\rho_2}$ reaches its infinite mass limit at $m_\pi/m_{\rho_2} \approx 0.7$ and then remains constant. Perhaps more importantly, the light quark mass result is also significantly lower for the $\rho_2$. There is no good reason to prefer the $\rho$ over the $\rho_2$ in presenting the Edinburgh plot, except that it is conventional. Clearly, until these differences are understood, one should plot both. The lines through the Wilson results represent fits to the masses and extrapolations to higher values of $\kappa$. A quark model calculation[42] of the hadron masses as a function of quark mass produces an Edinburgh plot that contains the "hump" seen with Wilson valence quarks. We see that the Kogut-Susskind $\rho$ results have a peak, but the shape is different from the Wilson case, whereas, the $\rho_2$ results have no peak where $m_\pi/m_\rho > 0.7$. Additional work at weaker coupling is necessary to verify that the two quark formulations give equivalent results. Studies with either quenched or dynamical quarks would be valuable.

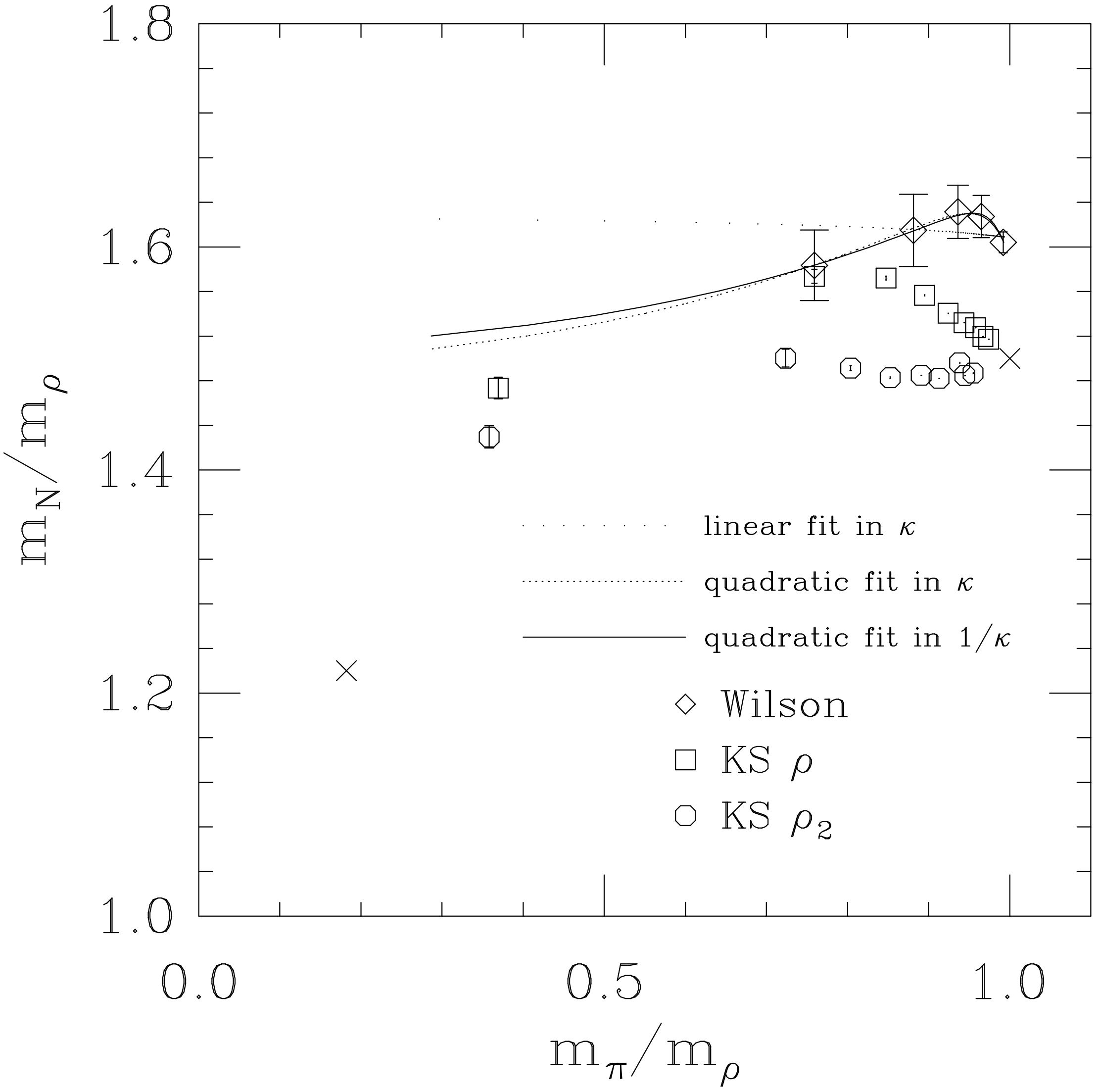

Fig. 15. Edinburgh plot. The octagons and squares are for Kogut-Susskind valence quarks, with the left most point corresponding to $m_v = m_d$. The diamonds are for Wilson valence quarks. Also shown are several extrapolations from the Wilson points toward the light quark limit. The quadratic fits in $\kappa$ or $\kappa^{-1}$ are much better than linear fits. The two crosses represent the physical and infinite quark mass values.

## VI. CONCLUSIONS

We have studied the hadron mass spectrum in full QCD with two flavors of Kogut-Susskind quarks at a fairly large lattice spacing. The large lattice spacing allows us to run with large physical volume and a reasonably small pion mass. Effects of the spatial size of the lattice were studied, as well as effects of the valence and sea quark masses.

We find small but statistically significant effects of the lattice size on the masses. Most of this effect comes between $N_s = 8$ and $N_s = 12$. These results should be useful in setting parameters for further simulations, as well as for comparison to models of the box size effects on QCD energies.

### Acknowledgements

We are very grateful to Sean Gavin and Sid Kahana for valuable discussions of finite density nuclear matter. The calculations reported in this paper were carried out on Intel iPSC/860 hypercubes located at the San Diego Supercomputer Center, the NASA Ames Research Center, and the Superconducting Supercollider Laboratory. We wish to thank all three centers for their support of our work. This research was supported in part by Department of Energy grants DE-2FG02–91ER–40628, DE-AC02–84ER–40125, DE-AC02–86ER–40253, DE-FG02–85ER–40213, DE-FG03–90ER–40546, DE-FG02–91ER–40661, and National Science Foundation grants NSF–PHY90–08482, NSF–PHY91–16964 and NSF–PHY-91–01853. Brookhaven National Laboratory is supported under DOE Contract DE–AC02–76CH00016.

## TABLE CAPTIONS

TABLE I. Summary of all runs showing spatial size $N_s$, molecular dynamics step size $\delta t$, time step between momentum refreshes $t$, total length of run, equilibration time and time considered equilibrated.

TABLE II. Particle masses from each run along with the $\chi^2$ of each fit, the number of degrees of freedom of the fit and the confidence level CL.

TABLE III. Energies of the particles carrying momentum, together with corresponding zero momentum fits.

TABLE IV. Pole positions and relative residues in the vector meson channel after taking account of mixing with two pion states.

TABLE V. Combined confidence level of all the fits with different spatial sizes for $am_q = 0.025$ for various values of $D_{min}$.

TABLE VI. Derivative of hadron masses w.r.t. valence or dynamical mass.

TABLE VII. Derivatives of the ratio of nucleon mass to $\rho$ or $\rho_2$ mass.

Table I

| $am_q = 0.025\ 6/g^2 = 5.445$ | | | | | |
|---|---|---|---|---|---|
| $N_s$ | $\delta t$ | $t$ | Length | Equilibration | Equilibrated |
| 8 | 0.02 | 1.0 | 1810 | 256 | 1554 |
| 8 | 0.02 | 1.0 | 2020 | 240 | 1780 |
| 10 | 0.02 | 1.0 | 1680 | 288 | 1392 |
| 12 | 0.02 | 1.0 | 1236 | 200 | 1036 |
| 16 | 0.02 | 1.0 | 2156 | 600 | 1556 |
| $am_q = 0.0125\ 6/g^2 = 5.415$ | | | | | |
| $N_s$ | $\delta t$ | $t$ | Length | Equilibration | Equilibrated |
| 12 | 0.01 | 0.5 | 1196 | 222 | 974 |
| 12 | 0.01 | 0.5 | 480 | 198 | 282 |
| 16 | 0.01 | 1.0 | 1428 | 208 | 1220 |

Table II

| Particle Masses for $am_q = 0.025$ | | | |
|---|---|---|---|
| Particle | Mass | $\chi^2$/d.o.f. | CL |
| $N_s = 8$ | | | |
| $\pi$ | 0.4529 (7) | 4.28/7 | 0.75 |
| $\rho$ | 0.949 (12) | 6.43/5 | 0.27 |
| $N$ | 1.456 (22) | 1.79/4 | 0.77 |
| $\pi_2$ | 0.779 (11) | 6.23/5 | 0.28 |
| $\rho_2$ | 0.994 (14) | 8.40/5 | 0.14 |
| $N_s = 10$ | | | |
| $\pi$ | 0.4500 (10) | 6.56/7 | 0.48 |
| $\rho$ | 0.938 (12) | 9.85/5 | 0.08 |
| $N$ | 1.408 (21) | 4.49/4 | 0.34 |
| $\pi_2$ | 0.763 (11) | 3.46/5 | 0.63 |
| $\rho_2$ | 0.953 (12) | 16.20/5 | 0.01 |
| $N_s = 12$ | | | |
| $\pi$ | 0.4496 (9) | 5.55/7 | 0.59 |
| $\rho$ | 0.909 (7) | 4.87/5 | 0.43 |
| $N$ | 1.351 (11) | 4.61/4 | 0.33 |
| $\pi_2$ | 0.754 (9) | 9.07/5 | 0.11 |
| $\rho_2$ | 0.950 (11) | 2.41/5 | 0.79 |
| $N_s = 16$ | | | |
| $\pi$ | 0.4488 (4) | 11.25/7 | 0.13 |
| $\rho$ | 0.918 (4) | 2.77/5 | 0.74 |
| $N$ | 1.375 (8) | 20.33/4 | $4 \times 10^{-4}$ |
| $\Delta$ | 1.43 (4) | 3.9/4 | 0.42 |
| $\pi_2$ | 0.759 (4) | 6.05/5 | 0.30 |
| $\rho_2$ | 0.954 (5) | 5.33/5 | 0.38 |

| Particle Masses for $am_q = 0.0125$ | | | |
|---|---|---|---|
| Particle | Mass | $\chi^2$/d.o.f. | CL |
| | $N_s = 12$ | | |
| $\pi$ | 0.3236 (4) | 4.37/7 | 0.74 |
| $\pi$ | 0.3235 (6) | 4.30/6 | 0.64 |
| $\rho$ | 0.894 (10) | 8.60/6 | 0.20 |
| $N$ | 1.284 (18) | 4.83/5 | 0.44 |
| $\pi_2$ | 0.676 (11) | 7.81/6 | 0.25 |
| $\rho_2$ | 0.904 (13) | 1.95/6 | 0.92 |
| | $N_s = 16$ | | |
| $\pi$ | 0.3244 (4) | 5.64/7 | 0.58 |
| $\pi$ | 0.3239 (5) | 5.06/6 | 0.54 |
| $\rho$ | 0.883 (6) | 7.87/6 | 0.25 |
| $N$ | 1.311 (10) | 3.05/5 | 0.69 |
| $\pi_2$ | 0.699 (7) | 3.68/6 | 0.72 |
| $\rho_2$ | 0.917 (8) | 9.88/6 | 0.13 |

Table III

| Particle | Momentum | Energy | Fit range | $\chi^2$/d.o.f. | CL |
|---|---|---|---|---|---|
| | | $am_q = 0.025$ | | | |
| Goldstone pi | 0 | 0.4478(3) | 8–11 | 0.56/2 | 0.76 |
| Goldstone pi | $2\pi/L$ | 0.5843(12) | 8–11 | 1.7/2 | 0.42 |
| $\gamma_5 \times \gamma_5\gamma_i$ pion | 0 | 0.764(3) | 5–12 | 5.2/4 | 0.27 |
| $\gamma_5 \times \gamma_5\gamma_i$ pion | $\parallel$ | 0.835(8) | 5–12 | 8.3/4 | 0.08 |
| $\gamma_5 \times \gamma_5\gamma_i$ pion | $\perp$ | 0.845(6) | 5–12 | 3.4/4 | 0.49 |
| $\gamma_i \times \gamma_i$ rho | 0 | 0.916(09) | 5–12 | 1.1/4 | 0.89 |
| $\gamma_i \times \gamma_i$ rho | $\parallel$ | 1.001(11) | 5–12 | 1.1/4 | 0.89 |
| $\gamma_i \times \gamma_i$ rho | $\perp$ | 0.995(13) | 5–12 | 0.70/4 | 0.95 |
| $\gamma_i \times \mathbf{1}$ rho | 0 | 0.947(5) | 3–10 | 6.5/4 | 0.16 |
| $\gamma_i \times \mathbf{1}$ rho | $\parallel$ | 1.044(8) | 3–10 | 2.1/4 | 0.72 |
| $\gamma_i \times \mathbf{1}$ rho | $\perp$ | 1.036(10) | 3–10 | 1.7/4 | 0.80 |
| | | $am_q = 0.0125$ | | | |
| Goldstone pi | 0 | 0.3242(19) | 7–12 | 8.8/4 | 0.07 |
| Goldstone pi | $2\pi/L$ | 0.5015(10) | 7–12 | 3.7/4 | 0.45 |
| $\gamma_5 \times \gamma_5\gamma_i$ pion | 0 | 0.710(3) | 4–11 | 15.0/4 | 0.004 |
| $\gamma_5 \times \gamma_5\gamma_i$ pion | $\parallel$ | 0.803(10) | 4–11 | 3.3/4 | 0.50 |
| $\gamma_5 \times \gamma_5\gamma_i$ pion | $\perp$ | 0.800(6) | 4–11 | 3.5/4 | 0.49 |
| $\gamma_i \times \gamma_i$ rho | 0 | 0.887(7) | 4–11 | 2.1/4 | 0.71 |
| $\gamma_i \times \gamma_i$ rho | $\parallel$ | 0.941(11) | 4–11 | 6.8/4 | 0.15 |
| $\gamma_i \times \gamma_i$ rho | $\perp$ | 0.968(12) | 4–11 | 6.1/4 | 0.19 |
| $\gamma_i \times \mathbf{1}$ rho | 0 | 0.942(7) | 3–10 | 1.9/4 | 0.97 |
| $\gamma_i \times \mathbf{1}$ rho | $\parallel$ | 1.000(16) | 3–10 | 7.9/4 | 0.10 |
| $\gamma_i \times \mathbf{1}$ rho | $\perp$ | 1.004(13) | 3–10 | 5.2/4 | 0.26 |

Table IV

| Polarization | Pole | Residue | Bare Pole | Representative momenta $(2\pi/L)$ |
|---|---|---|---|---|
| | | $\mu_L^2 = 1.0 \quad G^2 = 2.3 \quad \mu_\pi = 0.4478$ | | |
| $\parallel$ | 1.0228 | 0.505 | 1.0297 | $(0,0,0)+(0,0,1)$ |
| $\parallel$ | 1.0349 | 0.335 | 1.0366 | (bare rho) |
| $\parallel$ | 1.2824 | 0.001 | 1.2820 | $(0,1,0)+(0,-1,1)$ |
| $\parallel$ | 1.4459 | 0.001 | 1.4455 | $(0,0,2)+(0,0,-1)$ |
| $\parallel$ | 1.4866 | 0.000 | 1.4865 | $(1,1,0)+(-1,-1,1)$ |
| $\perp$ | 1.0265 | 0.838 | 1.0366 | (bare rho) |
| $\perp$ | 1.2830 | 0.003 | 1.2820 | $(0,1,0)+(0,-1,1)$ |
| $\perp$ | 1.4871 | 0.001 | 1.4865 | $(1,1,0)+(-1,-1,1)$ |
| | | $\mu_L^2 = 0.92 \quad G^2 = 2.3 \quad \mu_\pi = 0.3242$ | | |
| $\parallel$ | 0.8244 | 0.002 | 0.8248 | $(0,0,0)+(0,0,1)$ |
| $\parallel$ | 0.9933 | 0.841 | 0.9940 | (bare rho) |
| $\parallel$ | 1.1325 | 0.006 | 1.1316 | $(0,1,0)+(0,-1,1)$ |
| $\parallel$ | 1.3116 | 0.002 | 1.3110 | $(0,0,2)+(0,0,-1)$ |
| $\parallel$ | 1.3628 | 0.001 | 1.3626 | $(1,1,0)+(-1,-1,1)$ |
| $\perp$ | 0.9913 | 0.833 | 0.9940 | (bare rho) |
| $\perp$ | 1.1341 | 0.016 | 1.1316 | $(0,1,0)+(0,-1,1)$ |
| $\perp$ | 1.3637 | 0.003 | 1.3626 | $(1,1,0)+(-1,-1,1)$ |

Table V

| $D_{min}$ | $\sum \chi^2$ | d.o.f. | CCL |
|---|---|---|---|
| | | $\pi$ | |
| 1 | 29.9 | 32 | 0.572 |
| 2 | 27.6 | 28 | 0.483 |
| 3 | 23.2 | 24 | 0.510 |
| 4 | 17.8 | 20 | 0.598 |
| 7 | 19.0 | 16 | 0.268 |
| 8 | 14.5 | 12 | 0.271 |
| | | $\rho$ | |
| 3 | 75.8 | 24 | $3 \times 10^{-7}$ |
| 4 | 23.9 | 20 | 0.246 |
| 5 | 20.6 | 16 | 0.196 |
| 6 | 17.9 | 12 | 0.118 |
| | | Nucleon | |
| 3 | 71.3 | 20 | $1 \times 10^{-7}$ |
| 4 | 31.2 | 16 | 0.013 |
| 5 | 26.7 | 12 | 0.008 |
| | | $\pi_2$ | |
| 3 | 37.9 | 24 | 0.036 |
| 4 | 24.8 | 20 | 0.209 |
| 5 | 22.4 | 16 | 0.132 |
| | | $\rho_2$ | |
| 3 | 2821.7 | 24 | 0.000 |
| 4 | 32.3 | 20 | 0.040 |
| 5 | 28.3 | 16 | 0.029 |

Table VI

| Valence Mass Derivatives | | | | |
|---|---|---|---|---|
| Particle | $\frac{\partial m_h}{\partial m_v}$ | Fit range | $\chi^2$/d.o.f | CL |
| $\pi$ | 12.32(2) | 3–12 | 4.20/6 | 0.65 |
| $\pi$ | 12.34(3) | 4–12 | 1.25/5 | 0.94 |
| $\rho$ | 6.2(1.1) | 3–12 | 17.21/6 | 0.01 |
| $\rho$ | 5.1(1.6) | 4–12 | 6.62/5 | 0.25 |
| $\rho_2$ | 5.6(6) | 2–12 | 7.81/7 | 0.35 |
| $\rho_2$ | 7.0(1.0) | 4–12 | 7.05/5 | 0.22 |
| $N$ | 9.5(1.0) | 3–11 | 4.53/5 | 0.48 |
| $N$ | 9.5(4.2) | 4–11 | 2.28/4 | 0.68 |
| Dynamical Mass Derivatives | | | | |
| Particle | $\frac{\partial m_h}{\partial m_d}$ | Fit range | $\chi^2$/d.o.f | CL |
| $\pi$ | –0.11(16) | 2–12 | 4.82/7 | 0.68 |
| $\pi$ | –0.22(11) | 4–12 | 1.25/5 | 0.94 |
| $\rho$ | 5.9(4.0) | 3–12 | 17.21/6 | 0.01 |
| $\rho$ | 12.0(3.6) | 4–12 | 6.62/5 | 0.25 |
| $\rho_2$ | 9.9(1.3) | 2–12 | 7.81/7 | 0.35 |
| $\rho_2$ | 20.9(3.0) | 4–12 | 7.05/5 | 0.22 |
| $N$ | 8.3(3.9) | 3–9 | 3.36/3 | 0.34 |
| $N$ | 6.4(4.6) | 3–11 | 4.53/5 | 0.48 |
| $N$ | 15.2(13.8) | 4–11 | 2.28/4 | 0.68 |

Table VII

| | |
|---|---|
| $\frac{\partial(\frac{m_N}{m_\rho})}{\partial m_d}$ | –2.98(7.2) |
| $\frac{\partial(\frac{m_N}{m_{\rho_2}})}{\partial m_d}$ | –9.63(4.2) |
| $\frac{\partial(\frac{m_N}{m_\rho})}{\partial m_v}$ | 0.9(2.0) |
| $\frac{\partial(\frac{m_N}{m_{\rho_2}})}{\partial m_v}$ | –0.3(8) |